\documentclass{aastex61}
\usepackage{amsmath,amstext,amsbsy}
\usepackage[T1]{fontenc}
\usepackage{apjfonts} 
\usepackage{float}
\usepackage{placeins}
\usepackage{comment}
\usepackage{bm}
\usepackage{color, xcolor}	
\usepackage{hyperref}

\setcitestyle{notesep={ }}

\shorttitle{Super-Eddington Disk X-Ray reverberation }
\shortauthors{Thomsen et al.}

\defcitealias{Thomsen2019}{Paper~I}
\begin{document}

\title{Relativistic X-ray reverberation from super-Eddington accretion flow}

\author{Lars Lund Thomsen}
%\email{gfh112@connect.hku.hk}
%\author{Lars Lund Thomsen}
\affiliation{Department of Physics, University of Hong Kong, Pokfulam Road, Hong Kong, \hyperlink{blue}{gfh112@connect.hku.hk}}

\author{Lixin Dai}
\affiliation{Department of Physics, University of Hong Kong, Pokfulam Road, Hong Kong, \hyperlink{blue}{lixindai@hku.hk}}

\author{Erin Kara}
\affiliation{MIT Kavli Institute for Astrophysics and Space Research
Massachusetts Institute of Technology
77 Massachusetts Avenue, 37-241
Cambridge, MA 02139}

\author{Chris Reynolds}
\affiliation{Institute of Astronomy, University of Cambridge, Cambridge, CB3 0HA, UK}

\begin{abstract}
X-ray reverberation is a powerful technique which uses the echoes of the coronal emission reflected by a black hole accretion disk to map out the inner disk structure. While the theory of X-ray reverberation has been developed almost exclusively for standard thin disks, reverberation lags have recently been observed from likely super-Eddington accretion sources such as the jetted tidal disruption event Swift J1644+57. In this paper, we extend X-ray reverberation studies into the regime of super-Eddington accretion with a focus on investigating the lags in the fluorescent Fe K$\alpha$ line region. We find that the coronal photons are mostly reflected by the fast and optically thick winds launched from the super-Eddington accretion flow, and this 
funnel-like reflection geometry produces lag-frequency and lag-energy spectra with unique observable characteristics. The lag-frequency spectrum exhibits a step-function-like decline near the first zero-crossing point. As a result, the magnitude of the lag scales linearly with the black hole mass for a large parameter space, and the shape of the lag-energy spectrum remains almost independent of the choice of frequency bands. Not only can these features be used to distinguish super-Eddington accretion systems from sub-Eddington systems, but they are also key for constraining the  reflection geometry and extracting parameters from the observed lags.
When fitting the observed reverberation lag of Swift J1644+57 to our modeling, we find that the super-Eddington disk geometry is slightly preferred over the thin disk geometry, and we obtain a black hole mass of 5-6 million solar masses and a coronal height around 10 gravitational radii. 
\end{abstract}


\section{Introduction} \label{sec:intro}

The accretion onto supermassive black holes (SMBHs) plays a major role in shaping the evolution of the universe. The well-established $M_{\rm BH}-\sigma_{\rm bulge}$ relation between the mass of the SMBH ($M_{\rm BH}$) and the velocity dispersion of the stars in the bulge ($\sigma_{\rm bulge}$) indicates that the growth of the SMBHs are coupled to the growth of their host-galaxies
\citep[][]{Magorrian1998, Kormendy2013}. One key factor contributing to this scaling relation is believed to be the feedback in the forms of radiation, winds and relativistic jets produced by active galactic nuclei (AGNs) \citep{Silk1998, Heinz2006}.
The radiation produced from a black hole (BH) system is often expressed in terms of its Eddington luminosity $L_{\rm Edd}$:
\begin{equation}
    L_{\rm Edd} \approx 1.26 \times 10^{38} \Bigg( \frac{M
    _{\rm BH}}{M_{\odot}}\Bigg) \ {\rm erg \ s^{-1}}  
\end{equation}
where $M_\odot$ is the solar mass. The accretion rate corresponding to $L_{\rm Edd}$ is called the Eddington accretion rate $\dot{M}_{\rm Edd}$: 
\begin{equation}
 \dot{M}_{\rm Edd}= \frac{L_{\rm Edd}} {\eta \ c^{2}},
\end{equation}
where $c$ is the speed of light and $\eta$ is the radiative efficiency. For this study, we adopt the conventional value of 0.1.

The recent discovery of massive quasars with $10^{9-10} M_\odot$ at large redshifts of $z=6-8$ \citep{Mortlock2011, Wu2015, bandados2018} promotes the idea that super-Eddington accretion, in which gas is accreted onto a black hole faster than the Eddington accretion rate, likely plays an important role in the growth of quasars in the early universe. 
Many theoretical studies have been conducted to study super-Eddington accretion around black holes. In the conventional, semi-analytical models such as the `slim disk' or the `Polish doughnut' model, the accretion flow does not produce a wind \citep[see the review by][]{Abramowicz2013}.
However, recent general relativistic radiation magnetohydrodynamic (GRRMHD) simulations of super-Eddington disks \citep{Ohsuga2009,Jiang2014,McKinney2014, Sadowski2014} have unanimously shown that optically and geometrically thick winds are launched by the large radiation pressure in the disk. Such winds have anisotropic profiles with velocity decreasing and density increasing as the inclination deviates from the pole \citep{Sadowski16, Dai18}. 
The ultrafast outflow (UFO) component, with speeds of $v_r\approx {\rm few} \times 0.1c$, has been detected in several super-Eddington sources such as ultra-luminous X-ray sources (ULXs) \citep[e.g.,][]{Walton2016,Pinto2016, Pinto2017,Kosec2018} and tidal disruption events (TDEs) \citep[e.g.,][]{Kara2018}. Since powerful super-Eddington winds from AGNs can deeply impact their host galaxies \citep{King2003}, it is desirable to have a method to directly probe their geometry and energy.

X-ray reverberation is a technique developed to probe the structure of the BH accretion disk by analyzing the time-dependence and energy shifts of photons. This technique has been widely applied to study AGNs and X-ray binaries \citep[e.g.][]{Fabian89, Nowak1999, Reynold99, Zoghbi2012, Wilkins12,  Uttley2014, Cackett2014, Cackett2021}. 
These works usually adopt the  geometry of a point-like lamppost corona irradiating a geometrically thin, Keplerian disk.
In this standard picture, a very hot and compact corona is placed above the BH from where it emits non-thermal X-ray photons.
About half of the coronal photons freely escape to the observer, while the other half irradiate the cold accretion disk, giving rise to a reflection spectrum. 
The most prominent feature of the reflection spectrum is the Fe K$\alpha$ fluorescent lines, produced from the photoionization of an inner K-shell electron of Fe \citep{Fabian89, Matt93}. Due to general relativistic (GR) effects and the relativistic Doppler effect between the rotating disk and the observer, the Fe line profile becomes broadened and skewed. The spectral shape of the K$\alpha$ lines can be used to effectively probe the innermost disk geometry and constrain the BH spin in systems which have geometrically thin disks rotating with relativistic Keplerian speed and truncated at the innermost stable circular orbit (ISCO) \citep[e.g.,][]{Reynold99, Reynolds14, Cackett2014, Taylor18}. 
Moreover, temporal changes in the reflection spectrum are expected to lag behind the driving coronal continuum emission, because the reflected photons need to travel a longer distance to the observer and are gravitationally time delayed \citep{Shapiro64}. Therefore, resolving these lags can allow us to further constrain the BH parameters along with the coronal geometry and location \citep{Wilkins16, Cackett2014} as well as the disk vertical structure \citep{Taylor2018B}.

We propose that the well-established technique of X-ray reverberation can be extended to probe the structure of super-Eddington accretion flow. In this accretion regime, the coronal photons are expected to be reflected by the optically thick winds launched from the geometrically thick accretion flow. Therefore, the reflection and reverberation signals should in principle give us an insight into the wind geometry and kinematics. The best observational evidence of X-ray reverberation from super-Eddington accretion flow so far has come from the jetted TDE Swift J164449.3+573451 (Swift J1644 hereafter) \citep{Burrows2011, Bloom11, Levan2011, Zauderer2011}.  In this event, a star was disrupted by a SMBH and its debris accreted onto the SMBH at a super-Eddington accretion rate \citep{Rees1988, Evans1989, Guillochon2013}. The SMBH mass is estimated to be a few $\times 10^6~M_\odot$ constrained from the X-ray variability timescales. The very high isotropic X-ray luminosity can be explained using a relativistic jet launched from the magnetized accretion disk and beamed towards the observer \citep{Tchekhovskoy14}. A highly blueshifted Fe K$\alpha$ line and a relatively symmetric lag-energy spectrum in the Fe line region have been observed from this system \citep{Kara16}. Since the inner disk is likely aligned with the jet and therefore observed face-on, the most plausible explanation of the observed Fe line blueshift is that the observer is viewing down an optically thin funnel surrounded by the optically-thick winds launched by the super-Eddington disk, and the coronal photons are reflected by the fast-moving wind. 
Inspired by this discovery, we have previously conducted a theoretical study \citep[][; hereafter referred to as \citetalias{Thomsen2019}]{Thomsen2019} to investigate the characteristics of the Fe K$\alpha$ lines produced from super-Eddington accretion disks, in which we have demonstrated that such disks do produce Fe K$\alpha$ lines with signatures consistent with the one observed in Swift J1644.

In this paper, we extend the groundwork in \citetalias{Thomsen2019} to include the temporal response of the Fe K$\alpha$ fluorescent line, and further implement rigorous GR calculations from the corona to the reflection surface. The paper is structured as follows: In Sec. \ref{Sec:Irradiation}, we show the geometry, dynamics and ionization of the reflection surface for the super-Eddington accretion flow. We also calculate the emissivity profile under rigorous GR. In Sec. \ref{sec:Reverberation}, we present the 2D transfer function as well as the energy and frequency-dependent X-ray reverberation lags. Here we  compare the lags produced by super-Eddington disks and standard thin disks and show their characteristics are very different.
In Sec. \ref{Sec:Swift_Model}, we apply our model to the observed lags of Swift J1644 and fit various physical parameters. Lastly, in Sec. \ref{sec:discussion} we summarize our findings and discuss the prospect of using the X-ray reverberation technique to effectively extract information from super-Eddington accretion systems.

\section{Reflection geometry and emissivity profile} \label{Sec:Irradiation}

%In this way, we can also obtain the travel time and energy shift as photons travel from the corona to this reflection surface and calculate the emissivity profile. 

%This section is structured as the following. In Sec. \ref{sec:Disk_profile}, we calculate the geometry of the reflection surface and show the kinematics and ionization state of the gas along the reflection surface.  Next in Sec. \ref{sec:Emissivity}, we describe the general methodology in calculating the emissivity profile and show the results.  

\subsection{The geometry and kinematics of the reflection surface} \label{sec:Disk_profile}

Following \citetalias{Thomsen2019},
we use the super-Eddington disk profile from \citet{Dai18} simulated using the GRRMHD code HARMRAD \citep{McKinney2014, Mckinney15} for this study.
The simulated disk surrounds a SMBH with a mass of $M_{\rm BH}=5\times10^6~M_\odot$ and a fast spin with the dimensionless spin parameter $a=0.8$.
The disk has an average accretion rate of $\sim15~\dot{M}_{\rm Edd}$ and an average outflow rate of $\sim10~\dot{M}_{\rm Edd}$. 

\begin{figure}
\centering
\figurenum{1}
\begin{minipage}{0.48\textwidth}
\centering
\figurenum{1a}
\includegraphics[width=\linewidth]{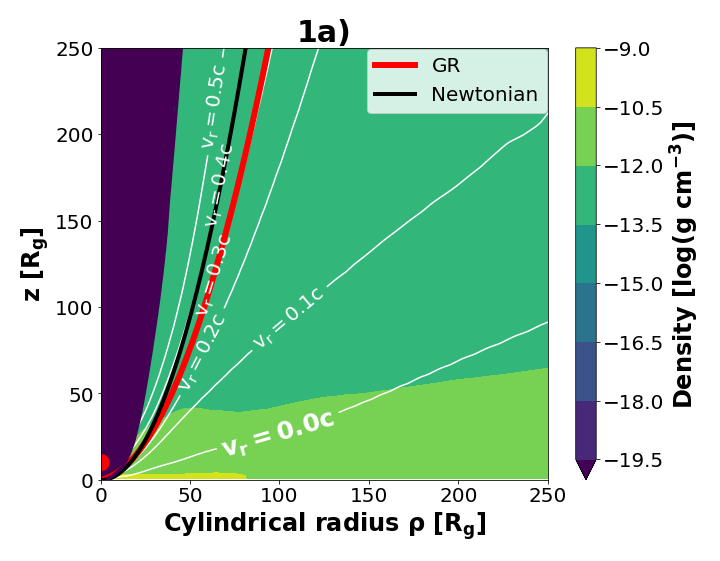}
\label{fig:Photosphere}
\end{minipage}
\begin{minipage}{0.48\textwidth}
\centering
\figurenum{1b}
\includegraphics[width=\linewidth]{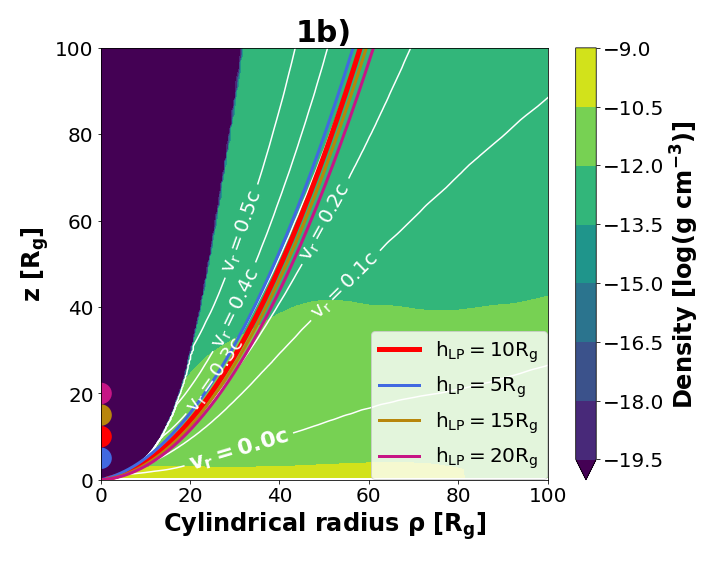}
\label{fig:Photosphere_hLP}
\end{minipage}
\caption{\textbf{The corona reflection geometry for the simulated super-Eddington disk.} The x-axis is the cylindrical radius $\rho$ and the y-axis is the vertical height $z$. The background color depicts the density profile of the disk. The jet is illustrated as the dark blue region around the pole which is assumed to be optically thin. 
The wind closer to the pole has a lower density and moves faster, with the white lines showing contours of constant $v_r$.
We place an artificial corona at some height $h_{\rm LP}$ and calculate the reflection surface of the coronal emission as the electron-scattering photosphere with a certain Thompson optical depth $\tau$. 
\textbf{1a. Comparison between the GR and the Newtonian reflection surface:} For $h_{\rm LP}=10R_g$ and $\tau=3$, the thick red curve shows the reflection surface calculated using rigorous GR ray-tracing, while the thick black curve shows the reflection surface calculated using Newtonian calculation. 
\textbf{1b. Reflection surfaces for different   $\pmb{h_{\rm LP}}$:} The red, thick curve is the same as the reflection surface in Fig. \ref{fig:Photosphere} with $h_{\rm LP} = 10 R_g$, while the thin curves show the GR reflection surfaces when the lamppost corona is placed at different $h_{\rm LP}$.
It can be seen that the coronal reflection surface is not sensitive to $h_{\rm LP}$. 
}
\end{figure}

For super-Eddington accretion flow, the coronal photons are reflected by the funnel wall, which is surrounded by optically thick winds. We illustrate this concept in Fig. \ref{fig:Photosphere} by showing the time and azimuthally-averaged disk profile.
The relativistic jet launched by the Blandford-Znajek process \citep{Blandford1977} is marked by the dark blue regions where the magnetic pressure dominates over the gas pressure. The density inside this region is ignored since the gas density at the base of the jet can be artificially boosted due to numerical reasons. The disk wind, launched by high radiation and magnetic pressure, is denser and slower towards the equator. This means an optically thin funnel is formed in the polar region, which has a half-opening angle around $10-15^\circ$ for the simulated disk we use. 
Therefore, an observer needs to look directly into the funnel to see the coronal emission and its reflection. If viewed at larger inclination angles, the X-ray photons are likely absorbed and reprocessed in the optically thick disk and wind.

We place an artificial lamppost corona at a few gravitational radii ($R_g$) above the black hole, which irradiates isotropically in its local (Minkowski) frame (see Sec. \ref{Appendix:Isotrpoic} for isotropic ray-tracing from a point source in a curved space-time).
While we recognize that the coronal photons should go through multiple scatterings inside the thick accretion flow, we take a simplified approach by assuming all the reflected photons originate from a single reflection surface in this work. This reflection surface is obtained by following the geodesics of photons emitted by the corona using our GR ray-tracing code (see Sec. \ref{sec:GR_Ray-tracing} for the description) until a certain optical depth $\tau$ is reached. Furthermore, we adopt the special relativistic correction to the optical depth $\tau$ \citep{Abramowicz1991}
to reflect that photons moving in the same direction as the gas flow should generally travel longer before being scattered:
\begin{equation}\label{eq:tau}
    \tau=\int \gamma \ \big(1-\beta \cos(\theta)\big) \ \kappa_{\rm es,0} \ \rho_0 \ ds,
\end{equation}
where $\kappa_{\rm es, 0} = 0.34 \ {\rm cm^2 \ g^{-1}}$ is the Thomson electron-scattering opacity (assuming solar abundance), $\rho_0$ is the rest-frame gas density, $\theta$ is the angle between the photon momentum and the gas velocity, and $ds$ is the length of the light path in the curved space-time. Here $\beta$ and $\gamma$ are the magnitude of the 3-velocity and the Lorentz factor between the frame of the gas and that of the Zero-Angular-Momentum-Observer (ZAMO) in the Kerr spacetime around the black hole. 
This rigorously relativistic approach used to calculate the reflection surface is different from the Newtonian approach used in \citetalias{Thomsen2019}.

We show in Fig. \ref{fig:Photosphere} the reflection surface obtained using this GR approach in comparison with the Newtonian one, both using
the $\tau=3$ photosphere from  a lamppost corona at the height of $h_{\rm LP}=10R_g$. 
To the first order, both surfaces lie in the optically thick, fast-moving winds with speeds at few$\times0.1c$.
The relativistic correction in Eq. \ref{eq:tau} causes the effective opacity to be smaller since the photon and gas generally move in the same direction. Therefore, the GR reflection surface lies at a larger inclination angle from the pole, with a slightly lower terminal speed compared to the Newtonian one.
We also vary the lamppost height between $5-20R_g$ and find that the reflection surface stays almost unchanged as seen in Fig. \ref{fig:Photosphere_hLP}.
Therefore, for simplicity, we adopt the reflection surface with $\tau=3$ calculated for lamppost corona at $h_{\rm LP}=10R_g$ throughout this study unless otherwise specified.

\begin{figure}
\centering
\figurenum{2}
\begin{minipage}{0.48\textwidth}
\centering
\figurenum{2a}
\includegraphics[width=\linewidth]{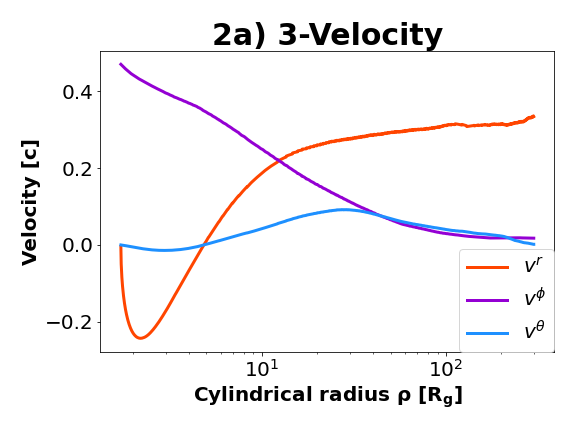}
\label{fig:vel}
\end{minipage}
\begin{minipage}{0.48\textwidth}
\centering
\figurenum{2b}
\includegraphics[width=\linewidth]{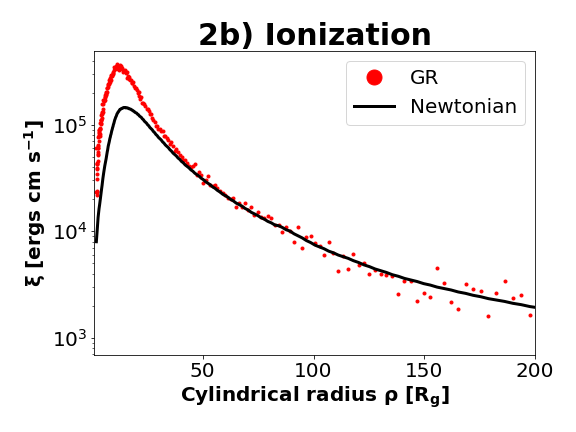} 
\label{fig:ionisation}
\end{minipage}
\caption{\textbf{2a. The equivalent Newtonian 3-velocity of the gas along the reflection surface}. The wind accelerates quickly and the out-flowing motion dominates over-rotation starting from $\rho \approx 10 \ R_{ g}$, after which the wind quickly reaches a terminal radial velocity of about 0.3c. 
\textbf{2b. The ionization parameter $\boldsymbol{\xi}$ along the reflection surface.}  By accounting for the GR effects (red dotted curve), $\xi$ is a few times larger in the inner disk region as compared to that obtained using Newtonian calculations (black thin curve). 
}
\end{figure}

Next, we check the kinematics of the reflection surface. Fig. \ref{fig:vel} shows the equivalent 3-velocity of the gas along the reflection surface. At small cylindrical radii $\rho<4 R_g$, the reflection surface lies within the disk inflow region, thus having negative radial velocities. 
Beyond $\rho\approx 10 R_g$, the reflection surface lies within fast winds and the radial motion dominates over its rotation.

Last, we investigate the X-ray ionization parameter of the gas, $\xi$, which determines whether Fe K$\alpha$ fluorescent lines can be produced, and if so, which lines (i.e., with rest-frame energy of 6.4, 6.7 or 6.97 keV) are produced \citep{Ross1999,Ballantyne2001, Garcia10}.
The ionization parameter is calculated as \citep{Reynolds97}:
\begin{equation}
    \xi(r)=\frac{4\pi F_{{\rm X-ray}}}{n(r)},
\end{equation}
where $n(r)$ is the gas number density at the reflection surface and $F_{{\rm X-ray}}$ is the hard X-ray flux over certain energy band \citep[e.g.,][uses 0.01-100keV]{Ballantyne2001}.
For this study, we assume that the efficiency of producing hard X-rays from accretion $\eta_X$ is about 1\%, similar to that in thin disks \citep{Reynold99}. Therefore, $F_{{\rm X-ray}}$ is scaled so that the total hard X-ray luminosity of the corona from the simulated disk is $L_{\rm X-ray} \approx \eta_x \ \dot{M} c^2 \approx 10^{44} \ {\rm erg \ s^{-1}}$.
As shown in Fig. \ref{fig:ionisation}, $\xi$ is enhanced by a factor of a few in the inner regions when including the GR effects (with details to be explained in the next section). 
$\xi$ is high in the super-Eddington case ($10^3 -10^5 \ \rm ergs\ cm\ s^{-1}$) compared to the thin disk scenarios ($\xi \lesssim100 \ \rm ergs\ cm\ s^{-1}$), since the reflection surface lies in the wind with relatively low densities. The high $\xi$ value favors the production of the Fe K$\alpha$ lines with larger rest-frame energies ($6.97$ and $6.7$ keV) over the $6.4$ keV line \citep{Ballantyne2001},  %Therefore, the rigorous GR calculation should be important for calculating the exact ionization of the super-Eddington accretion flow and determining the type of Fe lines produced,
although in this paper we assume the same rest-frame energy for the Fe K$\alpha$ lines produced throughout the entire reflection surface and  present the results in terms of the energy shift factor $g$ between the observed line energy and the emitted line energy.

\subsection{The emissivity profile} \label{sec:Emissivity}

The emissivity profile gives the strength of the coronal irradiating emission received at each annulus of the photosphere.
By following the photons from the corona to the reflection surface using GR ray-tracing, we automatically account for the light-bending effects, the energy shifts of the photons, and the Shapiro time delay. Other general relativistic effects, such as the general relativistic correction to the area of the reflection surface, need to be addressed separately.
Following \citet{Wilkins12, Dauser2013, Gonzalez2017}, the emissivity profile, $\epsilon (\rho)$, is given by: 
\begin{equation} \label{Eq:emissivity}
\epsilon(\rho) \propto \frac{N(\rho, d\rho)}{g_{\rm LP}^\Gamma \gamma A(\rho,d\rho)}.
\end{equation} 
Here $N(\rho, d\rho)$ is the number of coronal photons hitting an annulus of the photosphere at cylindrical radius $\rho$ with a width $d\rho$, and $A(\rho, d\rho)$ is the area of the annulus as seen by the ZAMO observer at the reflection surface. In order to calculate the area in a rigorous GR setting, we perform a Jacobian transformation from ($r(\rho), \theta(\rho), \phi$) to ($\rho, \phi$) and the details are presented in Sec. \ref{App:Area} following the approach in \citet[][]{Taylor18}. The Lorentz factor, $\gamma$, accounts for the length contraction of the area between the ZAMO observer and the rest-frame of the photosphere. Lastly, $g_{\rm LP}$ is the energy shift between the photons emitted from the corona ($E_{\rm LP}$) and the photons received by the reflection surface ($E_{\rm disk}$), which is defined as:
\begin{equation} \label{Eq:energy_shift_LP}
g_{\rm LP}=\frac{E_{\rm LP}}{E_{\rm disk}}= \frac{\big(p_\mu u^\mu\big)_{\rm LP}}{\big(p_\nu u^\nu\big)_{\rm disk}}.
\end{equation}
We propagate photons using the GR ray-tracing code and we calculate the energy as the product of the photon's four-momentum $p_\mu$ and the gas' four-velocity ($u^\mu$) evaluated at observation/emission. The power of the energy shift, $\Gamma$, in Eq. \ref{Eq:emissivity} arises from assuming a power-law emission from the corona with a photon index $\Gamma$ \citep{Dauser2013, Gonzalez2017}. We adopt the conventional value $\Gamma=2$ for this study, which is also appropriate for Swift J1644 \citep[][]{Burrows2011}. A detailed derivation of isotropic emission in a GR setting (how to transform from one frame to another) is presented in Sec. \ref{Appendix:Isotrpoic}.

\begin{figure}
    \centering
    \includegraphics[width=0.95\linewidth]{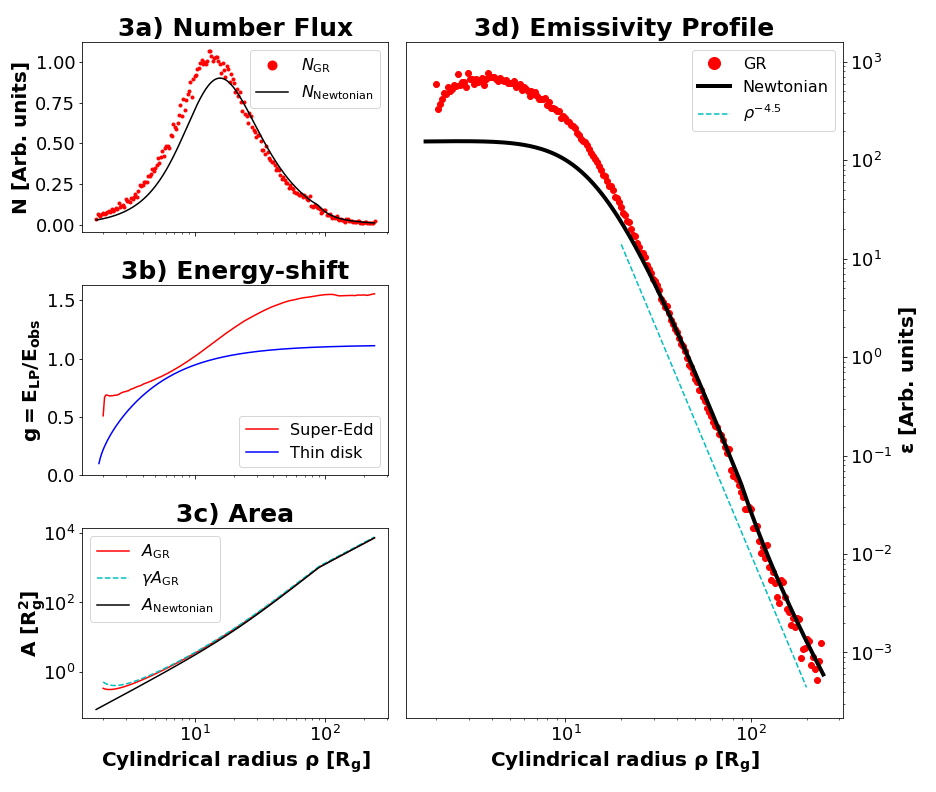}
    \figurenum{3d}
    \label{Fig:Emissivity_Emissivity}
    \figurenum{3a}
    \label{Fig:Emissivity_number}
    \figurenum{3b}
    \label{Fig:Emissivity_energy}
    \figurenum{3c}
    \label{Fig:Emissivity_area}
    \figurenum{3}
    \label{Fig:Emissivity}
    \caption{We show each component contributing towards the emissivity profile. 
    \textbf{3a) The number flux $\pmb{N}$:} The GR number flux is plotted together with the Newtonian one. We see the GR light-bending effects enhance the flux of photons in the inner region.
    \textbf{3b) The photon energy shift $\pmb{g_{\rm LP}}$:} We show the ratio between the energy of the emitted photons from the corona and the energy as seen by the gas elements in the reflection surface for both the super-Eddington and thin disks. $g_{\rm LP}<1$ means the photon as seen by the disk is blueshifted as compared to emission. 
    \textbf{3c) The area $\pmb{A}$:} We show 3 different areas: the classical Newtonian area (black solid), the proper area as seen by the ZAMO (cyan dashed), the relativistic proper area as seen by the reflection surface (red solid). One can see that the classical area substantially undervalues the area within $\rho=5R_g$ as compared to the two proper areas. 
    \textbf{3d) The emissivity profile $\pmb{\epsilon}$ of the super-Eddington accretion disk:} We show the GR emissivity profile (red dotted) (Eq. \ref{Eq:emissivity}) in comparison to the Newtonian one (black line). 
    The GR effects enhance the emission from the inner region. Approximately, the emissivity profile follows a power law $\propto \rho ^{-4.5}$ (cyan dashed) in the outer region and remains rather flat in the inner region.}
\end{figure}

We show in Fig. \ref{Fig:Emissivity} how each component above contributes to the total emissivity profile. 
In Fig. \ref{Fig:Emissivity_number}, we show the number flux $N(\rho, d\rho)$ of photons hitting the annulus calculated using the GR or Newtonian approach. The light-bending effect is clearly illustrated by the enhanced flux in the innermost $\rho<20R_g$ region. Next, in Fig. \ref{Fig:Emissivity_energy}, we show the photon energy shift $g_{\rm LP}$ from the corona to the reflection surface as a function of the cylindrical radius for both the super-Eddington disk and a standard thin disk. For both disks, the coronal photons experience gravitational blueshift when traveling down to the inner region of the reflection surface. For the super-Eddington disk geometry, the inner reflection surface is embedded in the disk region with a large inflow velocity, which gives a Doppler redshift for the coronal photons and therefore reduces the overall blueshift of the photons $g_{\rm LP}$. At the outer region of the super-Eddington accretion flow, the reflection surface lies in the outflow which moves away from the incoming coronal photons. Therefore, the energies of the coronal photons received by the reflection surface are Doppler redshifted.
In Fig. \ref{Fig:Emissivity_area} we show the relativistic proper area $A(\rho, d\rho)$ of the reflecting surface in comparison to the Newtonian area. In the innermost region, the relativistic area is amplified and the Lorentz factor further enhances this effect, which will slightly decrease the emissivity/irradiation in this region.

Putting these together, we obtain the GR emissivity $\epsilon(\rho)$ as shown in Fig. \ref{Fig:Emissivity_Emissivity}, which is compared to the Newtonian analytical emissivity profile calculated in \citetalias{Thomsen2019}. The GR emissivity profile is drastically enhanced in the inner region ($\rho\lesssim {\rm few} \times 10R_g$) mainly due to light-bending effects and the photon energy shift from the corona to the accretion flow. This increases the weight of the fluorescent lines produced in the inner disk regions. $\epsilon(\rho)$ can be fitted to a power law of $\epsilon \propto \rho ^{-4.5}$ far away, but stays roughly constant within $\rho \approx 10R_g$.

\section{Results} \label{sec:Reverberation}

\subsection{2D transfer Function} \label{sec:Reverberation:2dtrasnfer}
We first construct the 2D transfer function $\Psi(E,t)$ (also known as the impulse response function)  \citep{Reynold99, Uttley2014}. $\Psi(E,t)$ describes how the photons are emitted by an instantaneous coronal flare travel to different locations on the reflection surface and then get reflected towards the observer, by recording the unique observed energy ($E$), time delay ($t$), and intensity ($I_{\rm obs}$) of each reflected photon.
Here, $t$ measures the time difference between the arrival time of the reflected photons and the initial coronal flare. 
The energy shift of the reflected photon between emission and observation is calculated as: $g=E_{\rm obs}/E_{\rm emit}=(p_{\mu} u^{\mu})_{\rm obs}/(p_{\mu}  u^\mu)_{\rm emit}$.
The observed intensity of the photon goes as $I_{\rm obs}\propto g^4 $ (see Eq. \ref{Eq:energy_iobs} for derivation), which is weighted using the relativistic emissivity profile $\epsilon$ obtained in Sec. \ref{sec:Emissivity}.

\begin{figure}
\centering
\figurenum{4}
\begin{minipage}{0.32\textwidth}
\centering
\figurenum{4a}
\includegraphics[width=\linewidth]{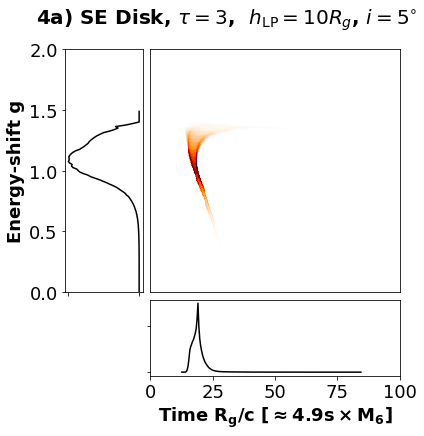}
\label{fig:2dtransfer_super-Edd_tau3}
\figurenum{4d}
\label{fig:2dtransfer_super-Edd_tau1}
\includegraphics[width=\linewidth]{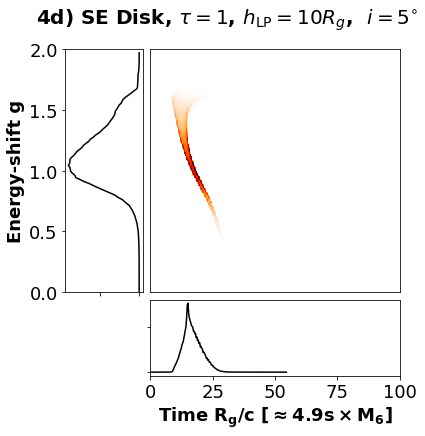}
\end{minipage}
\begin{minipage}{0.32\textwidth}
\centering
\figurenum{4b}
\label{fig:2dtransfer_super-Edd_LP20}
\includegraphics[width=\linewidth]{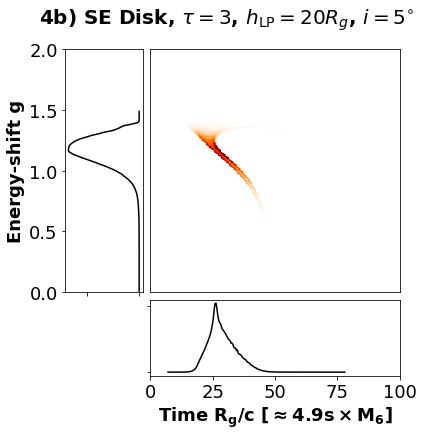}
\figurenum{4e}
\label{fig:2dtransfer_thin:low}
\includegraphics[width=\linewidth]{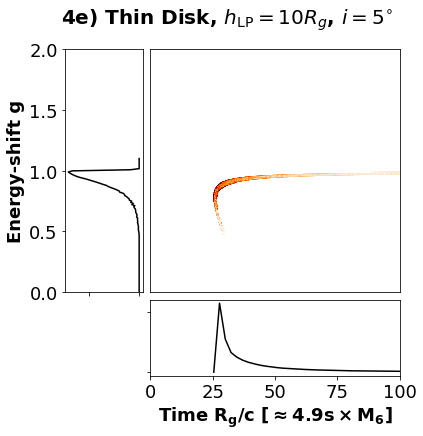}
\end{minipage}
\begin{minipage}{0.32\textwidth}
\centering
\figurenum{4c}
\label{fig:2dtransfer_super-Edd_LP5}
\includegraphics[width=\linewidth]{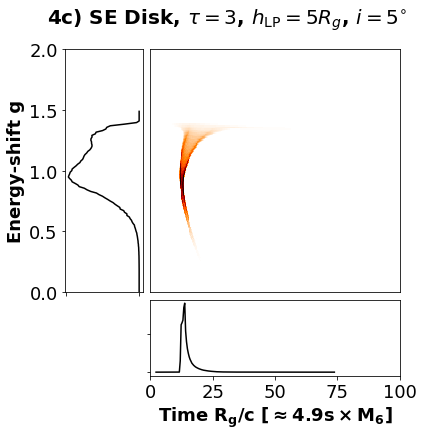}
\figurenum{4f}
\label{fig:2dtransfer_thin:high}
\includegraphics[width=\linewidth]{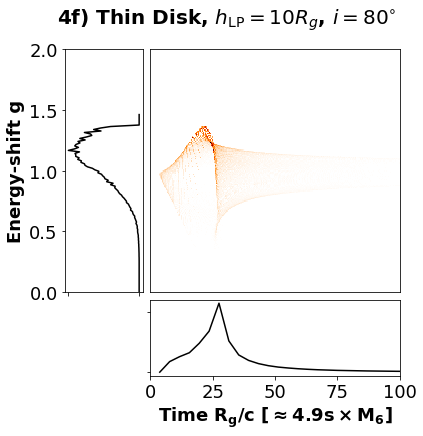}
\end{minipage}
\label{fig:2dtransfer}
\caption{\textbf{2D transfer functions for different super-Eddington (SE) and thin disk configurations.} 
The BH spin $a=0.8$ for all configurations. The x-axis shows the time delay of the reflected photons in gravitational units, which can be converted to real units as $R_g/c\approx 4.9s \times M_6$, with $M_6 \equiv M_{\rm BH}/10^6 M_\odot$. The energy shift is shown on the y-axis. Darker red color indicates stronger intensity $I_{\rm obs}$. The 1d response function, $\Psi(t)$, and the Fe K$\alpha$ line profile, $\Psi(E)$, are shown respectively in the bottom and left panels of each subfigure.
For SE disks, we fix the inclination angle to $i=5^\circ$, and vary the optical depth of the reflection surface $\tau$ and the lamppost height $h_{\rm LP}$ between $5-20R_g$ in Fig. \ref{fig:2dtransfer_super-Edd_tau3} - \ref{fig:2dtransfer_super-Edd_tau1}. As a comparison, for the thin disks, we have two inclination angles: face-on ($i=5^\circ$) (Fig. \ref{fig:2dtransfer_thin:low}) and edge-on ($i=80^\circ$) (Fig. \ref{fig:2dtransfer_thin:high}).}
\end{figure}

We show the 2D transfer functions of the super-Eddington disk and a standard razor-thin disk with the same BH spin $a=0.8$ for a direct comparison in Fig. \ref{fig:2dtransfer}. Throughout this study, unless otherwise specified, we adopt $h_{\rm LP}=10R_g$, a photosphere extending to $r_{\rm max}=1000R_g$, and an observer located at $r_{\rm obs}=5000R_g$. For the super-Eddington accretion flow, we put the observer at an inclination angle of $i=5^{^\circ}$, which is within the half-opening angle of the funnel $\theta_{\rm funnel}=15^{\circ}$. For the thin disk, the observer inclination is put at either $i=5^{^\circ}$ (low inclination) or $i=80^{\circ}$ (high inclination).
The 2D transfer functions are shown with time delays $t$ on the x-axis and energy shifts $g$ on the y-axis. The color indicates the intensity $I_{\rm obs}$ with the darkest regions being the strongest. By integrating $\Psi(E,t)$ over the energy, we obtain the 1d response function, $\Psi(t)$, which illustrates the intensity evolution of the reflected photons and is shown in the bottom panels of the subfigures. Likewise, the Fe K$\alpha$ line spectrum, $\Psi(E)$, is the time-averaged $\Psi(E,t)$ and is presented in the left-hand-side panels of the subfigures \footnote{\label{footnote:youtube}. For visualization purposes, we have produced videos showing how a coronal flare illuminates the accretion flow for: a super-Eddington disk \href{https://youtu.be/MYeRAN9MjB4}{https://youtu.be/MYeRAN9MjB4}; a low-inclination thin disk: \href{https://youtu.be/pVurM3540F4}{https://youtu.be/pVurM3540F4}; a high-inclination thin disk: \href{https://youtu.be/QfzyvJb-FGo}{https://youtu.be/pVurM3540F4}}.

We give a brief description of the behaviors of the 2D transfer functions, starting from the razor-thin disk one which has been extensively discussed in the literature \citep{Reynold99, Cackett2014, Ingram2019, Wilkins16, Wilkins20}. For a face-on thin disk (Fig. \ref{fig:2dtransfer_thin:low}), it takes longer for photons to travel to the outer part of the disk and get reflected towards the observer (upper branch), while photons reflected closer to the BH experience stronger Shapiro time delay and gravitational redshift (lower branch). Therefore, the first arriving photons are reflected at a few $R_g$ outside the inner edge of the disk. 
For an edge-on thin disk (Fig. \ref{fig:2dtransfer_thin:high}), the photons reflected by the outer part  of the disk on the side closest to the observer arrive first. 
Afterward, photons reflected by the inner disk are gradually received, where Doppler-shift largely broadens the width of the observed energy spectrum.
Then the maximum intensity is reached when photons from the innermost region of the disk arrive. And eventually the photons reflected by the outer disk on the other side reach the observer.

The super-Eddington disk $\Psi(E,t)$ for different $\tau$-surfaces and lamppost heights are shown in Fig. \ref{fig:2dtransfer_super-Edd_tau3}, \ref{fig:2dtransfer_super-Edd_LP20}, \ref{fig:2dtransfer_super-Edd_LP5} and \ref{fig:2dtransfer_super-Edd_tau1}.  
The overall shapes of these functions are similar to that of the low-inclination thin disk case. However, there are two notable differences: 1) The observed energy is largely blueshifted, due to the fast outflow speed of the funnel. 2) Most of the reflected photons are received within a short time delay range, which results from the narrow funnel reflection geometry in super-Eddington accretion.  
We can also see that varying the $\tau$-surface or $h_{\rm LP}$ does not change the first-order behavior of $\Psi(E,t)$. 
In general, decreasing the coronal height gives shorter time delays to the reflected photons. At the same time, the reflected spectrum gets less blueshifted and broader, since a lower corona height illuminates the inner, slower-moving part of the funnel more.

\begin{figure}
\centering
\centering
\figurenum{5}
\includegraphics[width=0.45\linewidth]{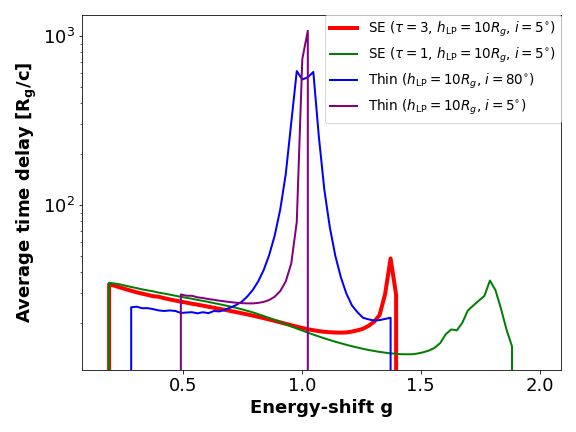}
\label{fig:energy-arrival}
\figurenum{5}
\caption{
\textbf{The average arrival time of photons as a function of energy shift for different disk configurations.} For thin disks, the longest average arrival time corresponds to $g\approx1$. For super-Eddington cases, it is found that the longest photon travel time corresponds to the regions with the largest red/blueshifts, i.e., the innermost/outermost regions of the reflection surface.
}
\end{figure}

In Fig. \ref{fig:energy-arrival}, we highlight the average time delay as a function of energy shift. For thin disks, we can see that the photons with the longest time delays have $g\approx1$, meaning these photons originate from the outer part of the disk. For super-Eddington disks, the photons with the longest time delays also have the largest redshifts (innermost region) or the largest blueshifts (outermost region). Also, the average time delays from super-Eddington disks are much smaller due to their reflection geometry.

\subsection{Frequency and Energy-dependent lags}  \label{sec:Reverberation:lag}

The variations of reflection dominated parts of the spectrum, such as the soft excess (0.1-1) keV band and the Fe K$\alpha$ (4-7) keV band, are expected to lag behind that of the continuum dominated part of the spectrum. Since the first detection of an Fe K-lags in AGNs by \citet{Zoghbi2012}, more than 20 AGNs with Fe K-lags have been observed and found to lag a few 100-1000s behind the continuum \citep[see][and references therein]{Kara2016c, Frederick2018, Vincentelli2020}. This lag timescale corresponds to the traveling time of photons reflected from the inner disk within a few gravitational radii ($t = \textrm{ few } \times R_g/c = \textrm{ few } \times 49 \frac{M_{\rm BH}}{10^7 M_\odot} s$). 
Since current X-ray observations usually do not have sufficient resolution to fully recover the 2D transfer function, one usually takes its Fourier transform to obtain the frequency or energy-dependent lags. This technique has been applied to perform X-ray reverberation studies of AGNs \citep{Zoghbi2013, Kara14, Cackett2021}.

We adopt the standard approach in \citet{Nowak1999, Cackett2014, Uttley2014} to calculate the frequency and energy dependent lags between the Fe line and the continuum. Assuming the coronal flare is a $\delta$-function impulse,  
the frequency dependent lags are calculated as \citep{Cackett2014}:
\begin{equation} \label{eqn:frequencylag}
    \phi(f)=\tan^{-1}\Big(\frac{R\times\Im{\big( \Tilde{\Psi}(f)}\big)}{1+R\times \Re{\big( \Tilde{\Psi}(f)}\big)} \Big), \ \ \ \ \textrm{and} \ \ \ T(f)=\frac{\phi(f)}{2\pi f}, 
\end{equation}
where $\Tilde{\Psi}(f)$ is the Fourier transform of the energy-averaged 1d response function $\Psi(t)$: $\Tilde{\Psi}(f)=\int_0 ^\infty \Psi(t) e^{-i2\pi ft}dt$, which is normalised ($\int \Psi(t) dt =1$).
The imaginary and real part of $\Tilde{\Psi}(f)$ are respectively $\Im{\big(\Tilde{\Psi}(f)}\big)$ and $\Re{\big(\Tilde{\Psi}(f)}\big)$. Here, $\phi(f)$ and $T(f)$ are the phase and time lag at Fourier frequency $f$ respectively.
The diluting factor, $R$, is defined as the ratio between the reflected flux and the driving coronal flux, which accounts for the dilution of the reflected Fe line flux by the continuum flux. A value of $R=1$ means an equal contribution of the continuum and reflected flux in the given energy bin. Furthermore, we assume $R$ to be constant across the entire Fe line, since we do not model the reflected flux strength relative to the continuum level. Therefore, the diluting factor decreases the overall lags by a factor $\sim R/(1+R)$.

The energy-dependent lags are calculated similarly. However, instead of averaging over all energies, we calculate the average lag in a given frequency band for each energy-bin of $\Psi(E,t)$. 
\begin{equation} \label{eq:energylag}
    \phi(E_j,f)=\tan^{-1}\Big(\frac{R\times\Im{\big(\Tilde{\Psi}(E_j,f)}\big)}{1+R\times \Re{\big(\Tilde{\Psi}(E_j,f)}}\big) \Big), \ \ \ \ \textrm{and} \ \ \ T(E_j)=\textrm{mean}\Big(\frac{\phi(E_j,f)}{2\pi f}\Big).
\end{equation}
Here, $\Tilde{\Psi}(E_j,f)=\int_0 ^\infty \Psi(E_j,t) e^{-i2\pi ft}dt$ and $\Psi(E_j,t)$ is normalised, such the energy where the line peaks, $E_0$, is $\int \Psi(E_0,t) dt=1$ \citep{Cackett2014}. $T(E_j)$ is the energy dependent lag averaged over the chosen frequency band.
We perform zero-padding when calculating the Discrete Fourier Transform (DFT) to increase the frequency resolution of the transform by adding 0s at the end of the time vector - leading to a more closely sampled Fourier frequency vector.

\subsubsection{Frequency dependence of lags} \label{Sec:Freq-lag}

\begin{figure}
\centering
\figurenum{6}
\begin{minipage}{0.45\textwidth}
\centering
\figurenum{6a}
\includegraphics[width=\linewidth]{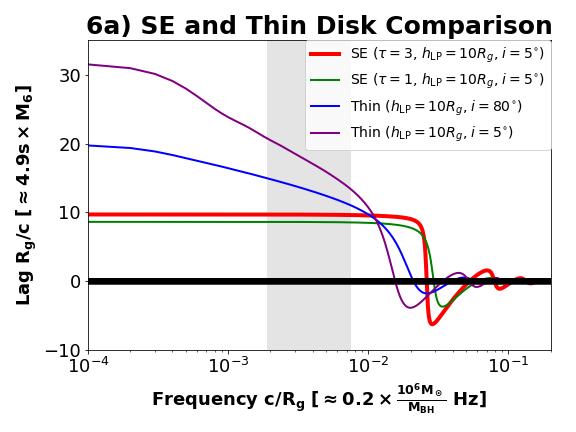}
\label{fig:freq-lag_compare}
\end{minipage}
\begin{minipage}{0.45\textwidth}
\centering
\figurenum{6b}
\includegraphics[width=\linewidth]{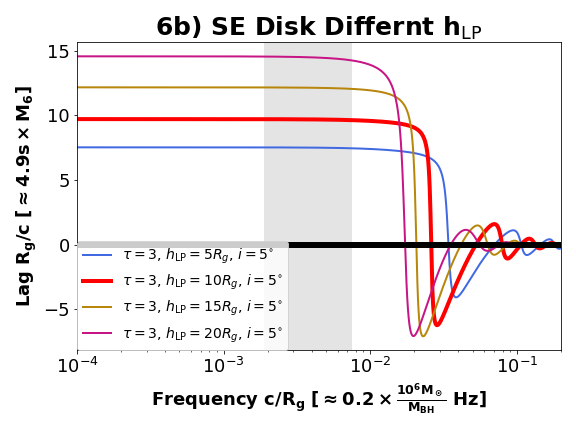}
\label{fig:freq-lag_LP}
\end{minipage}
\caption{\textbf{Lag-frequency spectra}:  The frequency and time units are expressed in natural units. The BH spin $a=0.8$, and the reflection surface is cut at $R_{\rm out}=1000R_g$ in all cases. The shaded region is the low-frequency band used to calculate the energy-dependent lags in Sec. \ref{Sec:Energy-lag}.
\textbf{6a:} Comparison between SE and thin disk cases (see labels for the configurations).  $h_{\rm LP}$ is fixed at $10R_g$. 
\textbf{6b:} Comparison between different $h_{\rm LP}$ for the SE disk.
}
\label{fig:freq-lag}
\end{figure}

We calculate the lag-frequency spectra of the super-Eddington accretion and thin disks using Eq. \ref{eqn:frequencylag} and show the results in Fig. \ref{fig:freq-lag}.
The x-axis, the Fourier frequency $f$, corresponds to the emissions with characteristic time delays shorter than $t=1/2f$. Therefore, the low frequencies probe all reflected emissions, while the high frequencies probe only the emissions with the shortest time delays. The y-axis, the lag, indicates the average time delay of photons. The lag stays constant at very low frequencies until a frequency $f_o\approx 1/2t_{\rm max}$, where $t_{\rm max}$ is the maximum time delay between the direct coronal and reflected photons as seen by a faraway observer. 
As frequencies increases above $f_0$, the photons with longer time delays are gradually phase-wrapped out. Therefore, the average time delay and, hence, the overall lag decreases with increasing frequencies. This trend lasts until the phase wraps completely and the lags turn negative, which is also known as the first zero lag crossing \citep[see][Sec. 4 for a more detailed description]{Uttley2014}. Beyond the first zero lag crossing, the lags become negative and then have a few oscillations at higher frequencies due to an artifact from Fourier analysis \citep{Cackett2014, Uttley2014}. 
%Since the phase of the lag is assumed to be in the range $-\pi$ to $\pi$, we are unable to distinguish whether the phase shift has a lag of $-\pi/2$ or a lead of $3\pi/2$. Therefore, seeing a negative lag at high frequencies does not necessarily mean that the continuum lags behind the Fe reflection line at those frequencies.  

We compare the lag-frequency spectra produced from the super-Eddington and thin disk structures in Fig. \ref{fig:freq-lag_compare}.
The most prominent difference is that the lags of the super-Eddington disks change abruptly from being nearly constant to rapidly decaying, whereas the thin disk lags decay more gradually.  
This is because for super-Eddington disks, as shown in Fig.  \ref{fig:2dtransfer_super-Edd_tau3} to \ref{fig:2dtransfer_super-Edd_tau1}, photons being reflected by the narrow funnel have a very short time delay within $t_{\rm max} = $ few$\times 10 R_g/c$. Therefore, we have $f_0 \approx 1/2t_{\rm max}\approx 1/(\rm{few}\times 10~R_g/c) \approx ~ \rm{few}\times 10^{-2}~c/R_g$. For the thin disks, $t_{\rm max}$ depends on the inclination and the outer radius of the disk. Since we set the disks with a radius of $1000 R_g$, their phase wrapping starts from $f_o \approx 1/(\rm{few} \times 1000~R_g/c)\sim \rm{few} \times 10^{-4}~c/R_g$. The rate of the lag decay reflects how concentrated the photon time delays are. The super-Eddington cases have a compact 1d response function $\Psi(t)$, so their frequency dependent lags have a sharp drop at $f_0$. 
The thin disks, however, have more gradually decreasing lags because their 1d response functions have a large spread. The fine structure of the lag spectra, such as the faster decay around $\rm{few} \times 10^{-2}~c/R_g$ for thin disks, indicate that most photons can travel to the observer within ${\rm few} \times 10 R_g/c$.

The first zero lag crossing (at frequency $f_{\rm zl}$) is roughly set by the peak of $\Psi(t)$. 
A lower $f_{\rm zl}$ indicates more photons return at a later time. 
Therefore, we can see in Fig. \ref{fig:freq-lag_LP} that increasing $h_{\rm LP}$ leads to longer lags as well as lower $f_{\rm zl}$. We can approximate $f_{\rm zl} \propto c/(4h_{\rm LP})$.
One might notice that if we plot the frequency-dependent lag spectrum in physical units, then $h_{\rm LP}$ and $M_{\rm BH}$ are somewhat degenerate, since increasing $M_{\rm BH}$ will also linearly increase the physical time lags and decrease the frequency of $f_{\rm zl}$. 
However, this degeneracy can be broken by calculating the energy-dependent lags which we will show in Sec. \ref{Sec:Energy-lag}.

%Overall, we find that the a super-Eddington disk will produce a frequency dependent lag spectrum which has a abrupt drop from a nearly constant lag to the zero lag crossing. This behaviour is very different from the the razor thin disk spectrum, which decays more slowly in their lag spectrum. For X-ray observations covering a cadence of $10^{-4}-10^{-3}$ Hz, we expect to see this constant part of the lag. Therefore, we can use this as an observable feature to distinguish the super-Eddington funnel geometry from the thin disk geometry. As this constant lag scales linearly with the coronal height $T(f) \approx {\rm few}\times h_{\rm LP}/c$ and $h_{\rm LP}$ is usually assumed to be a few $R_g$, it gives us some clue about the BH mass, although there is still a degeneracy between the coronal height and BH mass. 

It is worth noting that \citet[][]{Wilkins2013} have shown that if the corona has any physical extent the frequency-lag decay will become even smoother for thin disks, which might also apply to the super-Eddington case.
Also, the very slow decay of the thin disk lag-frequency spectra from constant lags to $f_{zl}$ as discussed above results from using the standard $\alpha$-disk model with no vertical dimension.
As a comparison, \citet[][]{Taylor2018B} have performed X-ray reverberation studies using realistic thin disk models with geometric thickness. They have found that as the disk becomes thicker as a result of increasing accretion rate, the disk lag-frequency spectra decays at a faster rate, which is qualitatively consistent with our results. Therefore, we expect that the lag-frequency spectrum evolves from slow decay to very sharp decay as the accretion rate increases from just below the Eddington accretion rate to the super-Eddington accretion rate.

\subsubsection{Energy dependence of lags} \label{Sec:Energy-lag}

In this subsection, we investigate the energy dependence of the Fe K lags. The lag-energy spectrum depends on the frequency band used to carry out the calculations in Eq. \ref{eq:energylag}. Throughout this section, we mainly assume a frequency band $\Delta f_{\rm XMM}=(0.9-3.6)\times 10^{-4} Hz = (1.9-7.6) \times 10^{-3}~\big(M_{\rm BH} /(5\times 10^6 M_\odot)\big)~c/R_g$, which is a typical low frequency range used in observational analysis \citep[e.g.,][]{Kara14}. For our fiducial $M_{\rm BH}=5\times 10^6 M_\odot$, this frequency band is marked as the shaded gray region in Fig. \ref{fig:freq-lag}, where it can be seen that in this frequency range the lag-frequency spectrum is still in the main decay phase.

\begin{figure}
\centering
\centering
\begin{minipage}{0.45\textwidth}
\figurenum{7a}
\includegraphics[width=\linewidth]{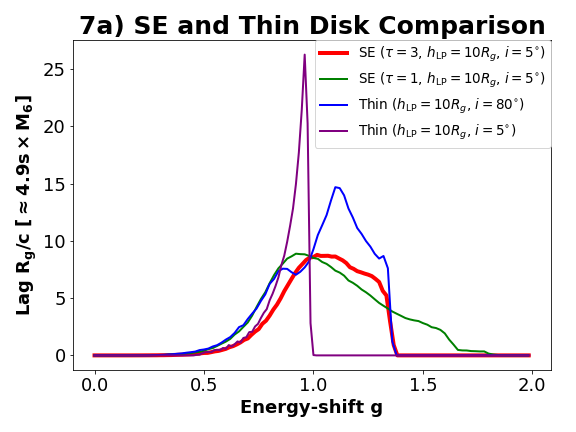}
\label{fig:energy-lag_compare}
\end{minipage}
\begin{minipage}{0.45\textwidth}
\includegraphics[width=\linewidth]{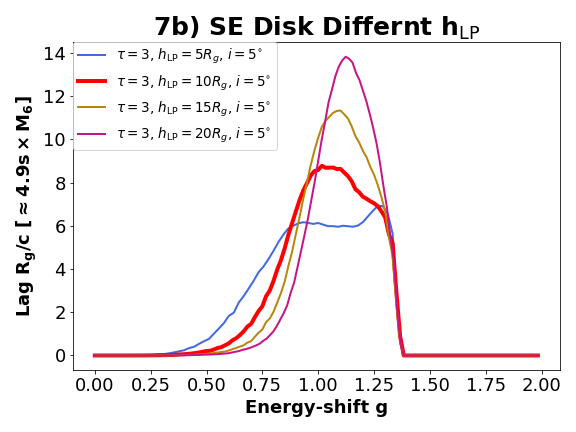}
\figurenum{7b}
\label{fig:energy-lag_LP}
\figurenum{7c}
\end{minipage}
\figurenum{7}
\caption{ \textbf{Lag-energy spectra} in the observed frequency band $\Delta f_{\rm XMM}=(0.9-3.6)\times 10^{-4} Hz = (1.9-7.6) \times 10^{-3} c/R_g$. 
\textbf{7a:} Comparison between the SE disk and thin disk cases (see legends).  $L_{\rm LP} = 10 R_g$.
\textbf{7b:} Comparison between different $L_{\rm LP}$ for the SE disk.
}
\label{fig:energy-lag}
\end{figure}

\begin{figure}
\centering
\begin{minipage}{0.32\textwidth}
\centering
\figurenum{8a}
\includegraphics[width=\linewidth]{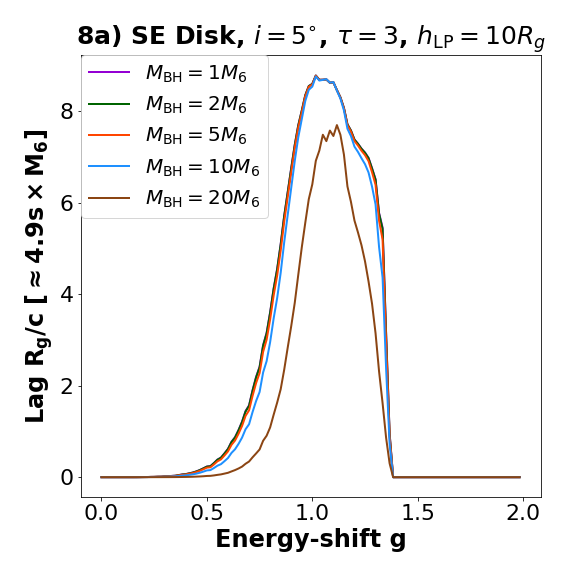}
\label{fig:energy-lag_MBH_super-Edd}
\end{minipage}
\begin{minipage}{0.32\textwidth}
\centering
\figurenum{8b}
\includegraphics[width=\linewidth]{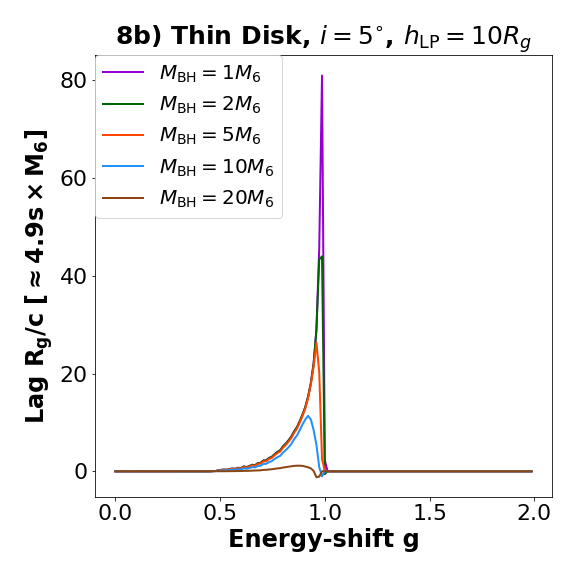}
\label{fig:energy-lag_MBH_thin_5}
\end{minipage}
\begin{minipage}{0.32\textwidth}
\centering
\figurenum{8c}
\includegraphics[width=\linewidth]{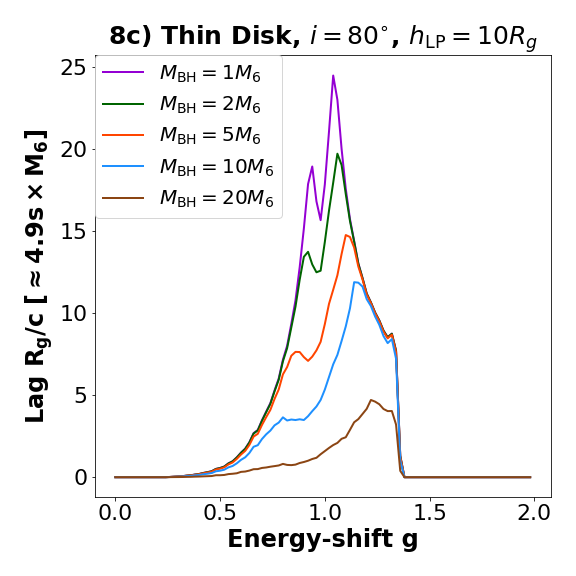}
\label{fig:energy-lag_MBH_thin_80}
\end{minipage}
\figurenum{8}
\caption{We show how the lag-energy spectrum changes with the BH mass in the frequency band $\Delta f_{\rm XMM}=(0.9-3.6)\times 10^{-4} $ Hz,
for a few representative $M_{\rm BH}=(1,2,5,10,20) \times M_6$, where $M_6=10^6 M_{\rm \odot}$. Note that we show the lags in units of $R_g/c$, so its magnitude in physical units should be multiplied with $M_{\rm BH}$.
\textbf{8a:} For the SE disk, the profile of the lag-energy spectrum remains almost constant as the $M_{\rm BH}$ increases all the way up to $2\times 10^7 M_{\odot}$. \textbf{8b \& 8c:} For a thin disk, its lag-energy profile changes substantially with $M_{\rm BH}$. Also the lag magnitude does not exactly scale linearly with $M_{\rm BH}$.}
\label{fig:energy-lag_mass}
\end{figure}

We first compare the lag-energy spectra produced from thin and super-Eddington disks in Fig. \ref{fig:energy-lag_compare}. It has been shown previously that for thin disks the shape of the Fe K lag-energy spectrum roughly mimics that of the Fe K line spectrum \citep{Kara13, Cackett2014}. 
Here we find that the same holds for the super-Eddington case. 
The lag-energy spectra from super-Eddington disks are more blueshifted than their thin disk counterparts and generally have less skewed profiles. For the high-inclination thin disk case, the lag-energy spectrum is even more skewed to the right than its spectral counterpart as seen in Fig. \ref{fig:2dtransfer_thin:high}. 
This is caused by emissions with the longest time delays which originate from the outer disk in the opposite side of the observer (with $g\approx1$) have been omitted due to the onset of phase-wrapping (see Fig. \ref{fig:energy-arrival}). 
In Fig. \ref{fig:energy-lag_compare} and \ref{fig:energy-lag_LP}, we also vary the $\tau$-surface or the coronal height for the super-Eddington disk case. One can see that despite the variance in these parameters, the lag-energy spectrum of the super-Eddington case stays largely blueshifted and rather symmetric in shape. 
We also find that the profile of the lag-energy spectrum from the super-Eddington geometry is quite sensitive to the lamppost height.
By lowering the coronal height, the irradiation of the inner disk increases, which introduces more gravitational redshift to the lag-energy spectrum and broadens the overall profile.

Next, we investigate how the lag-energy spectrum depends on the BH mass. Here we assume that all super-Eddington disks, regardless of their $M_{\rm BH}$ or other parameters, all have the same reflection geometry as the one we use. The frequency band in natural units as shown in Fig. \ref{fig:freq-lag} is correlated with the BH mass as $f= c/R_g\propto 1/M_{\rm BH}$. Therefore, increasing $M_{\rm BH}$ equivalently shifts the chosen frequency band closer to the first zero lag crossing, and its effect on the lag-energy spectrum is shown Fig. \ref{fig:energy-lag_mass}. Overall, for the super-Eddington disk (Fig. \ref{fig:energy-lag_MBH_super-Edd}), 
the shape of the lag-energy spectrum is almost independent of $M_{\rm BH}$
, especially for $M_{\rm BH}\lesssim 10^7 M_\odot$. 
The reason is that for this $M_{\rm BH}$ range we are sampling from the almost constant part of the lag-frequency spectrum at frequencies lower than $f_{\rm zl}$. 
Only for the largest $M_{\rm BH}=2\times 10^7 M_\odot$, the lag-energy profile becomes slightly narrower. This is a result of the frequency range starts to approach the first zero lag crossing with increasing $M_{\rm BH}$, the photons with the longest time delays are phase-wrapped out, which come from either the base of the funnel with long Shapiro time delays and the largest redshift or the outermost reflection surface with the longest travel paths and the largest blueshift. 
As a comparison, the shapes of the lag-energy spectra of thin disks change substantially with $M_{\rm BH}$ (Fig. \ref{fig:energy-lag_MBH_thin_80} and \ref{fig:energy-lag_MBH_thin_5}), because the onset of phase-wrapping occurs at very low frequencies for thin disks. For a low-inclination thin disk, the photons with the longest time delays come from the edge of the disk where the energy shifts are minimal and $g\approx 1$. Therefore, as $M_{\rm BH}$ increases, the central peak component at $g\approx 1$ decreases. For a high-inclination thin disk, the photons with the longest time delays come from the outer disk opposite to the observer. Therefore, as $M_{\rm BH}$ increases, the lag-energy spectrum transits from a more Newtonian-like double horn profile to a more relativistically skewed and broadened profile. 
Another important difference we find is that the magnitude of the lag (in units of $R_g/c$) remains constant in the super-Eddington case, while that of a thin disk can vary with $M_{\rm BH}$. For example, for the high inclination thin disk, the lags (in units of $R_g/c$) decreases by a factor of 5 as the $M_{\rm BH}$ increases from $10^6 M_\odot$ to  $2\times 10^7 M_\odot$.

\begin{figure}
\centering
\begin{minipage}{0.32\textwidth}
\centering
\figurenum{9a}
\includegraphics[width=\linewidth]{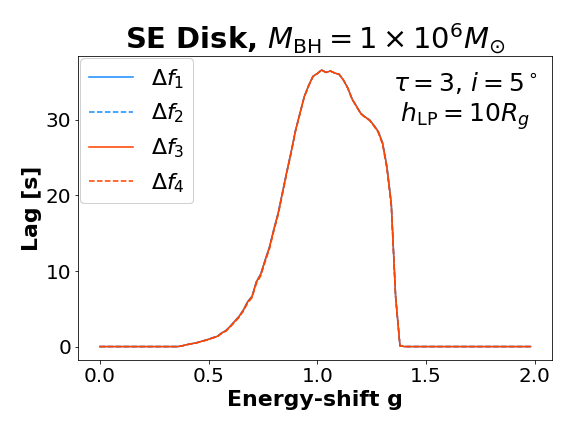}
\includegraphics[width=\linewidth]{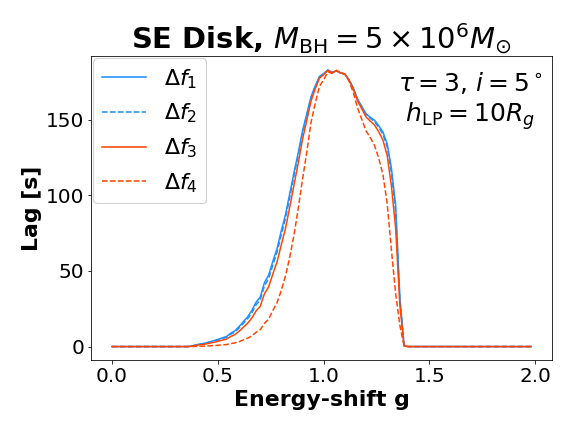}
\includegraphics[width=\linewidth]{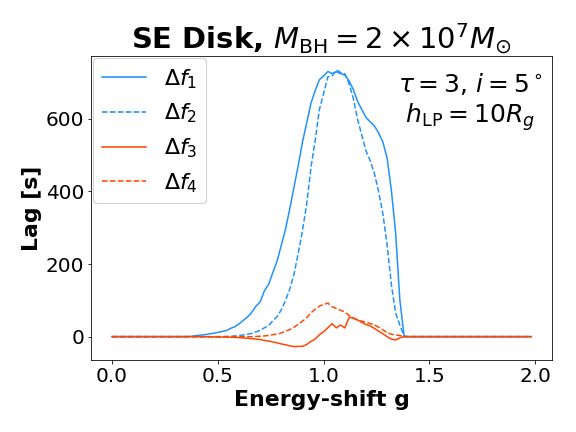}
\end{minipage}
\begin{minipage}{0.32\textwidth}
\centering
\figurenum{9b}
\includegraphics[width=\linewidth]{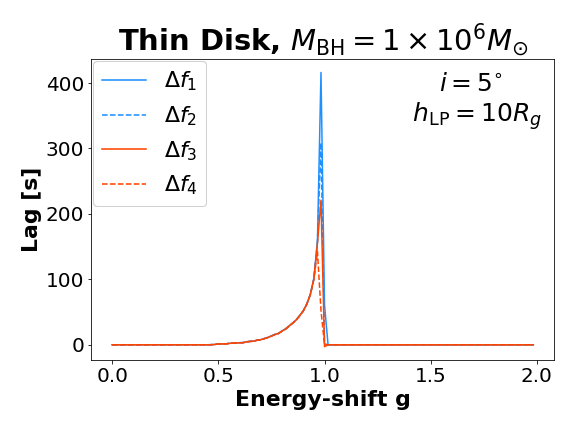}
\includegraphics[width=\linewidth]{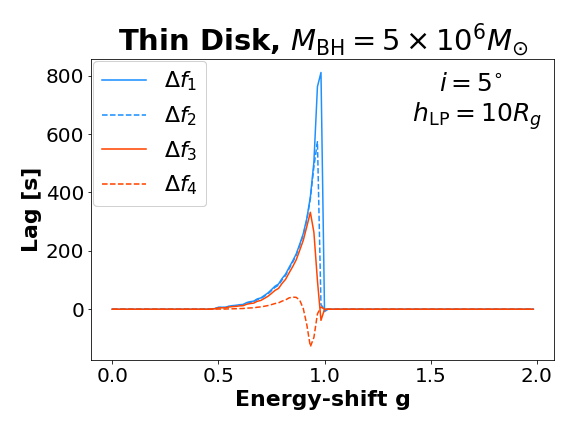}
\includegraphics[width=\linewidth]{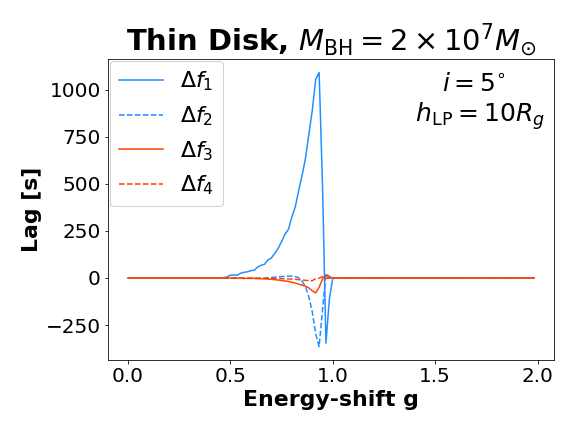}
\end{minipage}
\begin{minipage}{0.32\textwidth}
\centering
\figurenum{9c}
\includegraphics[width=\linewidth]{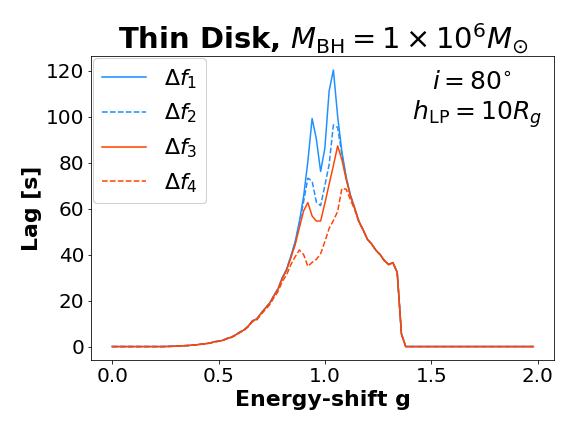}
\includegraphics[width=\linewidth]{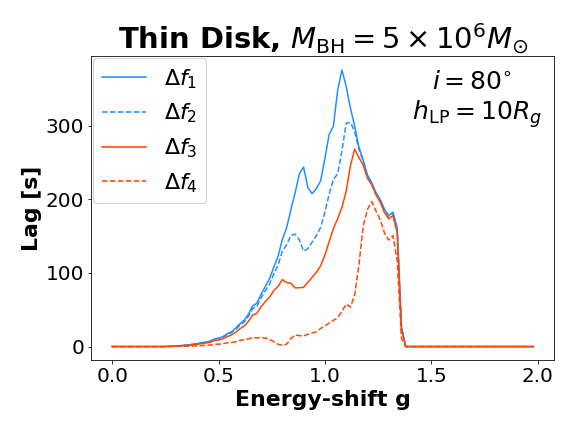}
\includegraphics[width=\linewidth]{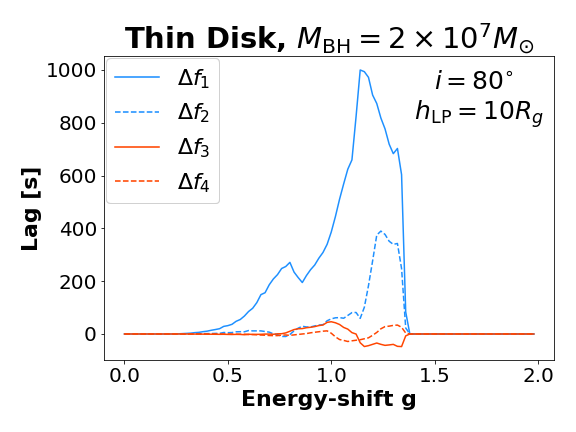}
\end{minipage}
\figurenum{9}
\caption{We show explicitly how the lag-energy spectrum depends on the exact frequency band. Here we use narrower frequency bands.
The blue solid/dotted curves represent the narrow low-frequency bands of
$\mathbf{\Delta f_1}=(0.8-1.6)\times 10^{-4} Hz=(0.34-0.67)\times 10^{-3} M_6 \ c/R_g$ and $\mathbf{\Delta f_2}=(1.6-3.2)\times 10^{-4} Hz=(0.67-1.34)\times 10^{-3} M_6 \ c/R_g$, where $M_6=M_{\rm BH}/10^6 M_\odot$.
The red solid/dotted curves are for the narrow high-frequency bands $\mathbf{\Delta f_3}=(2-6)\times 10^{-4} Hz=(0.84-2.52)\times 10^{-3}M_6\  c/R_g$ and $\mathbf{\Delta f_4}=(6-10)\times 10^{-4} Hz=(2.52-4.20)\times 10^{-3} M_6 \ c/R_g$. 
The three columns (left to right) represent different disk configurations: a) the SE disk, b) a low-inclination thin disk, and c) a high-inclination thin disk. Each row has a different $M_{\rm BH} = (1,5,20)\times 10^6 M_{\odot}$. 
The most notable distinction between the two disk geometries is that the lag-energy spectrum of the SE disk remains almost the same despite the change in the frequency band until $M_{\rm BH}$ becomes sufficiently large. 
}
\label{fig:energy-lag_narrow}
\end{figure}

Last, we show how  the lag-energy spectrum depends on the exact frequency band in Fig. \ref{fig:energy-lag_narrow}. The possible frequency bands used for analyzing data are usually set by the cadence of the telescope and the brightness of the object. Here we choose 4 narrow frequency bands.
The first two narrow frequency bands adopted are $\Delta f_1=(0.8-1.6)\times 10^{-4} $ Hz and $\Delta f_2=(1.6-3.2)\times 10^{-4} $ Hz, which are also inspired by the observations such as \citet{Kara14} (solid and dotted blue lines in Fig. \ref{fig:energy-lag_narrow}). The last two high frequency bands are $\Delta f_3=(2-6)\times 10^{-4} $ Hz and $\Delta f_4=(6-10)\times 10^{-4} $ Hz, which arise from splitting the observed frequency band of Swift J1644 \citep{Kara16} into two (solid and dotted red lines). The three columns in Fig. \ref{fig:energy-lag_narrow} represent the different disk configurations: the super-Eddington disk, a low inclination thin disk and a high inclination thin disk. The three rows represent different $M_{\rm BH} = (1, 5, 20)\times10^6 M_\odot$.
Here we see the scenario is similar to when we change $M_{\rm BH}$. For the super-Eddington disk, the shape of the lag-energy spectrum is almost the same for the four chosen frequency bands as long as $M_{\rm BH}<2\times 10^7 M_\odot$. On the contrary, the thin disks have their lag-energy spectra varying with all frequency bands. 
Therefore, we propose that if the lag-energy spectrum of the Fe K line remains constant between two close, narrow frequency bands, it means the 1d response function is very compact, which is a sign of a narrow funnel reflection. 
The exact change in the lag-energy profile we expect to see between the two narrow frequency bands depends on the $M_{\rm BH}$, $h_{\rm LP}$, frequency band, inclination angle, etc. To be conservative, for the $M_{\rm BH}$ and frequency bands ranges in Fig. \ref{fig:energy-lag_narrow}, if we see less than 5-10$\%$ difference in the lag magnitude and no morphological change in the lag profile between two narrow frequency bands, it is likely a super-Eddington accreting source.

\section{Application onto the super-Eddington TDE Swift J1644} \label{Sec:Swift_Model}

In this section, we model the observed X-ray reverberation lags of the jetted TDE Swift J1644. 
Using the 25ks XMM-Newton observation over multiple energy bands from 0.3 to 10 keV, \citet{Kara16} discovered that the jetted TDE showed signs of a strongly blueshifted Fe K line.
Furthermore, they found that the 5.5-8 keV band lagged around 100s behind the continuum dominated regions at 4-5 keV and 9-13 keV in the galaxy's rest-frame frequency band ($\Delta f_{\rm swift}=(2-10)\times 10^{-4} (1+z) $ Hz with the host galaxy at redshift $z=0.354$), and in the 5.5-8 keV band a Fe K lag was preferred over a constant lag with $>99.9\%$ confidence interval. The observed lag-energy spectrum was shown to be consistent with a largely blueshifted reflection component.

\begin{figure}
\centering
\begin{minipage}[t]{0.32\textwidth}
\centering{\Large \textbf{\ \ \ \ \ \ Lag-Energy}}\vspace{0.2 cm}
\includegraphics[width=\linewidth]{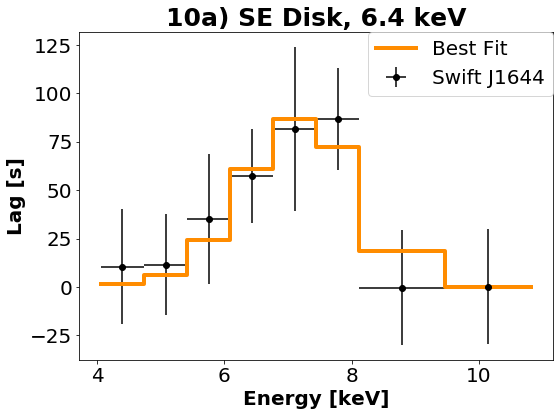}
\figurenum{10a}
\label{fig:energy-lag_swift_64}
\includegraphics[width=\linewidth]{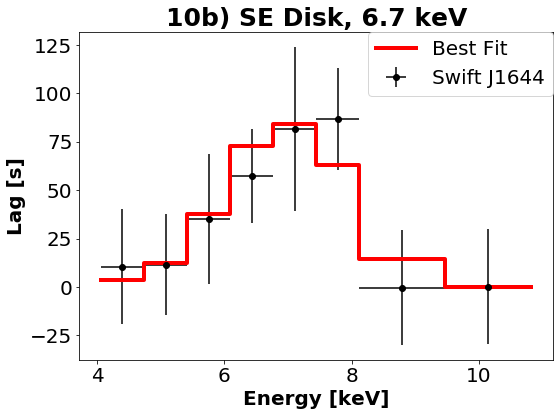}
\figurenum{10b}
\label{fig:energy-lag_swift_67}
\includegraphics[width=\linewidth]{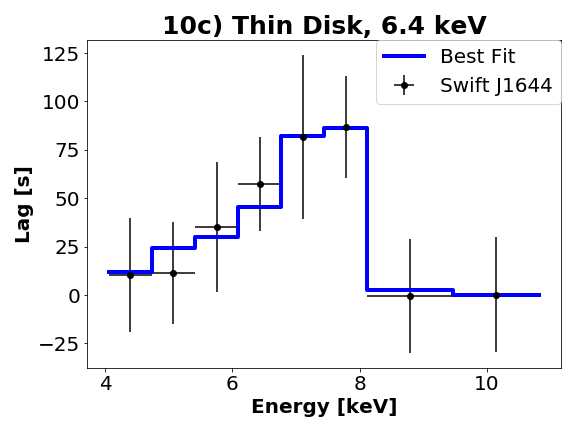}
\figurenum{10c}
\label{fig:energy-lag_swift_thin_64}
\includegraphics[width=\linewidth]{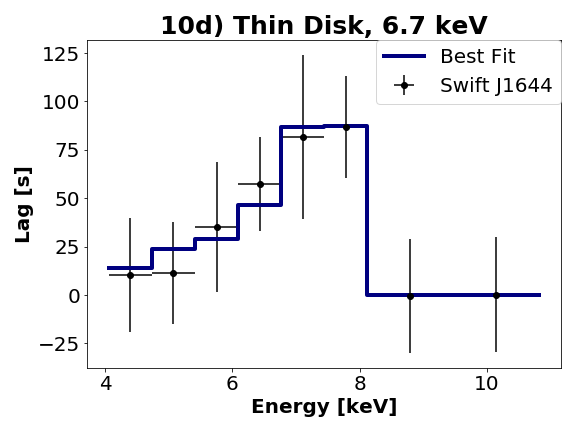}
\figurenum{10d}
\label{fig:energy-lag_swift_thin_67}
\end{minipage}
\begin{minipage}[t]{0.32\textwidth}
\centering{\Large \textbf{\ \ \ \ \ \ Lag-Frequency}}\vspace{0.2 cm}
\includegraphics[width=\linewidth]{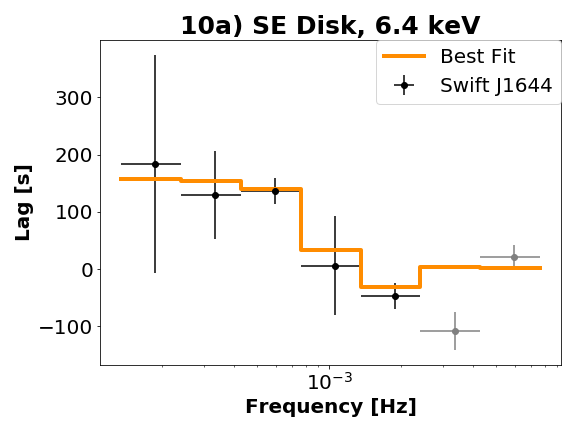}
\includegraphics[width=\linewidth]{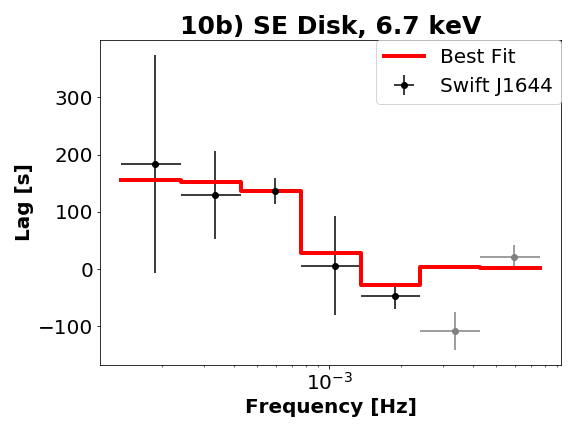}
\includegraphics[width=\linewidth]{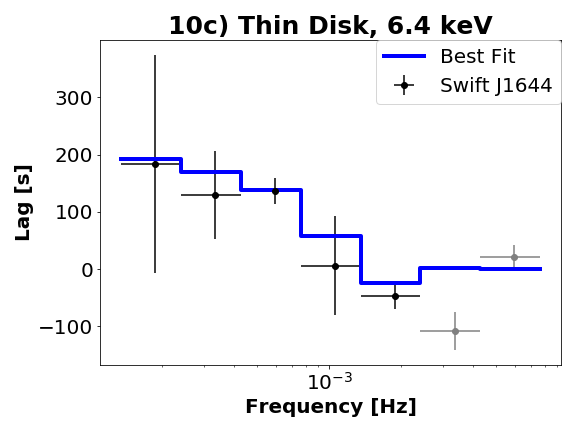}
\includegraphics[width=\linewidth]{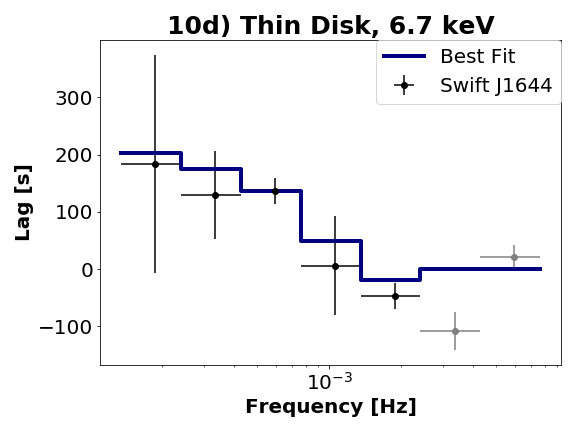}
\end{minipage}
\begin{minipage}[t]{0.32\textwidth}
    \centering{\Large \textbf{Best Fit Parameters}} \\[0.5cm]
     \framebox{
        \parbox{ 0.8\columnwidth}{ \centering \normalsize $M_{BH}=5.0\times 10^6 M_\odot $  \\  $h_{LP}=13R_g$ \\ $R=0.57$ \\ $\chi^2=13.55$ \\ $\chi_{\nu}^2=1.13$ \\ $\mathrm{BIC}=21.68$ \\[0.5cm] }}
        \\
    \vspace{1.1cm}
    \framebox{
        \parbox{ 0.8\columnwidth}{ \centering \normalsize $M_{BH}=5.9\times 10^6 M_\odot $  \\  $h_{LP}=10R_g$ \\ $R=0.54$ \\ $\chi^2=14.06$ \\ $\chi_{\nu}^2=1.17 $ \\ $\mathrm{BIC}=22.19$ \\[0.5 cm] }}
        \\
    \vspace{1.1cm}
    \framebox{
        \parbox{ 0.8\columnwidth}{ \centering \normalsize $M_{BH}=2.8\times 10^6 M_\odot $  \\  $h_{LP}=15R_g$ \\ $R=0.74$ \\ $i=70^\circ$ \\ $\chi^2=13.76$ \\ $\chi_{\nu}^2=1.25 $ \\ $\mathrm{BIC}=24.59$\\[0.1 cm] }}
        \\
    \vspace{1.05cm}
    \framebox{
        \parbox{ 0.8\columnwidth}{ \centering \normalsize $M_{BH}=3.5\times 10^6 M_\odot $  \\  $h_{LP}=11R_g$ \\ $R=0.77$ \\ $i=65^\circ$ \\ $\chi^2=13.74$ \\ $\chi_{\nu}^2=1.25 $ \\ $\mathrm{BIC}=24.58$ 
        \\[0.1 cm]} } 
\end{minipage}
\figurenum{10}
\label{fig:Best_fit}
\caption{We apply the MCMC algorithm to obtain the best fits to the observed frequency and energy-dependent lags of Swift J1644 using both the SE accretion disk model and thin disk model, and assume the Fe K line has rest-frame energy of either 6.4 or 6.7 keV. We vary the mass of the BH ($M_{\rm BH}=[10^5-10^8]~M_\odot$), the height of the corona ($h_{\rm LP}=[5-100]~R_g$), and the dilution factor ($R=[0.01, 2]$). For the SE disk model, the observer inclination angle $i$ is fixed at $5^\circ $, while for the thin disk model $i$ is an extra free parameter in the range of $[35^\circ - 80^\circ]$. The best fit parameters are listed on the right side of each row together with the chi-square, reduced chi-square and BIC values. The Swift J1644 observed lag spectra are plotted as the black dots with error bars, and the best-fit modeled lag spectra are plotted using colored lines. All observational points are within 1$\sigma$ from all model predictions except the last two high-frequency lags points in Fig. 9b (colored gray). At these high frequencies, it is increasingly difficult to disentangle the lag contribution due to a loss of coherence \citep[see][Section 2]{Uttley2014} and the Poisson Noise further reduce the S/N in this regime, so the two data points are not trustworthy.
The best fit parameters across all configurations indicate a black hole mass of $M_{\rm BH}=(2-6) \times 10^6 M_\odot$ with the lamppost corona located at $h_{\rm LP}=10-15 R_g$ above the black hole. The SE models are slightly preferred over the thin disk models. Also, see Fig. \ref{fig:swift_MCMC} for the 1D and 2D probability functions of the MCMC fit, the most probable parameters and the uncertainty.}
\label{fig:swift_best_fit}
\end{figure}

\begin{figure}
\centering
\begin{minipage}[t]{0.475\textwidth}
\figurenum{11a}
\label{fig:MCMC_64}
\includegraphics[width=\linewidth]{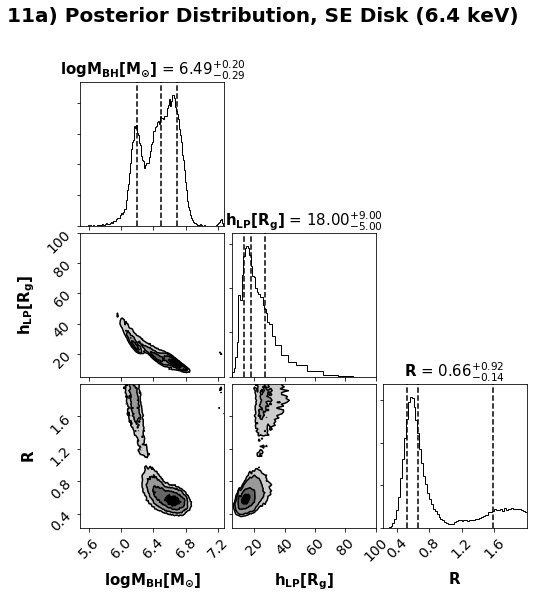}
\figurenum{11c}
\includegraphics[width=\linewidth]{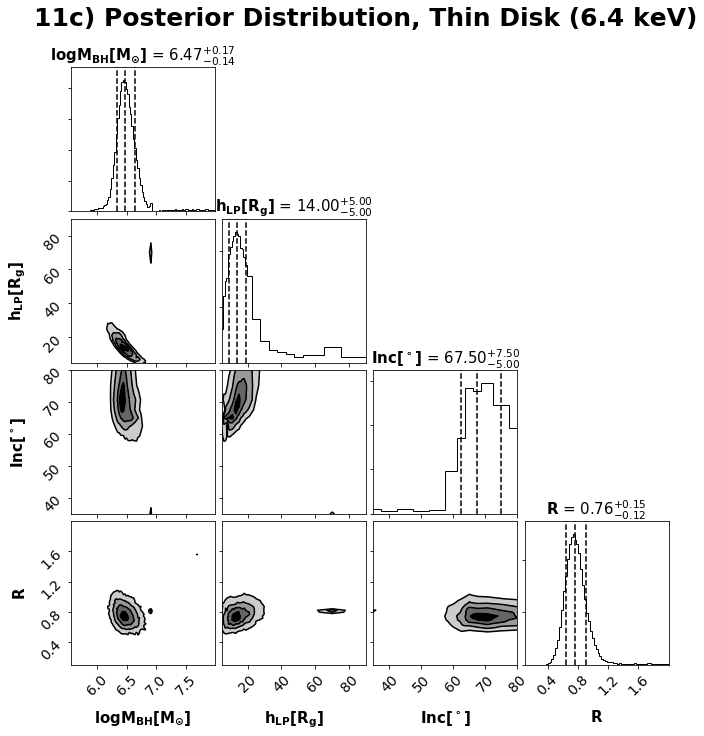}
\label{fig:MCMC_thin_64}
\end{minipage}
\begin{minipage}[t]{0.475\textwidth}
\figurenum{11b}
\label{fig:MCMC_67}
\includegraphics[width=\linewidth]{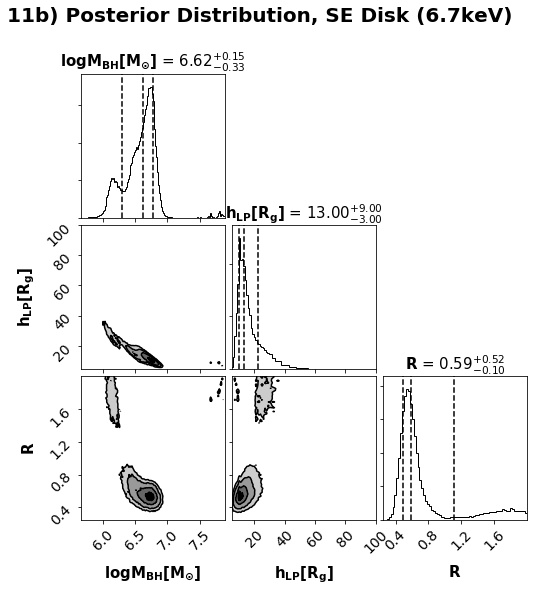}
\figurenum{11d}
\label{fig:MCMC_thin_67}
\includegraphics[width=\linewidth]{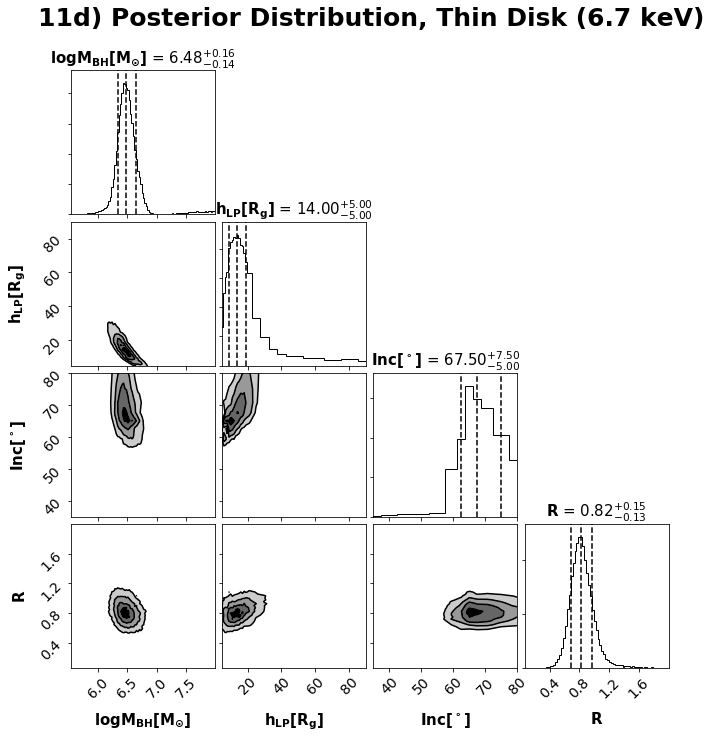}
\end{minipage}
\figurenum{11}
\label{fig:swift_MCMC}
\caption{ The 1d histograms located on the top of each column show the posterior distribution for each parameter from the MCMC algorithm. Note that the values shown are the most probable values of the parameters, which are different from best-fit parameters giving the lowest $\chi^2$ presented in Fig. \ref{fig:swift_best_fit}. The vertical dashed lines mark the 16th, 50th, and 84th percentile of probability. Also, for each parameter pair, we show the joint 2D histogram of the posterior distribution. The contours correspond to the 16th, 50th, and 84th percentile. 
}
\end{figure}

We apply our theoretical calculation of X-ray reverberation to model the observed lags of Swift J1644. 
We use a Markov Chain Monte Carlo (MCMC) module called \textit{emcee} in Python  \citep{emcee13} to draw random sets of parameters ($M_{\rm BH},h_{\rm LP}, R$) and simultaneously fit the observed lag-energy spectrum and the observed lag-frequency spectrum (the latter is calculated between the continuum dominated energies at $(3-4)\times (1+z)$ keV and the reflection dominated energies at $(5-6)\times(1+z)$ keV for this study). This MCMC method, based on a Bayesian framework, calculates the log-likelihood function  $-\chi^2/2$, where $\chi^2$ is the chi-square. The best fit with parameters minimizing the log-likelihood function is shown in Fig. \ref{fig:Best_fit}, where we also show the 1D and 2D posterior probability distributions of the parameters using the Python package Corner \citep{corner} in Fig. \ref{fig:swift_MCMC}.
When performing the fitting routine, we vary the mass of the BH $M_{\rm BH}=[10^5-10^8]~M_\odot$, the height of the corona $h_{\rm LP}=[5-100]~R_g$ and the dilution factor $R=[0.01, 2]$. For the super-Eddington disk, we assume that the reflection surface always stays the same as our fiducial one, and the inclination angle is fixed at $5^\circ$. For the thin disk, we vary the inclination angle between $i=[35-80]^\circ$ to fit a blueshifted reflection component. 
The rest-frame energy of the Fe K$\alpha$ line is tested for both $6.4$ keV (usually assumed for thin disks) and $6.7$ keV (expected for super-Eddington disks due to the high ionization level).

The best fits for the super-Eddington disk model with Fe K$\alpha$ rest-frame energies of $6.4$ and $6.7$ keV are shown in Fig. \ref{fig:energy-lag_swift_64} and \ref{fig:energy-lag_swift_67} respectively. We find that the super-Eddington disk model can simultaneously explain the observed energy and frequency-dependent Fe K lags, with a reduced chi-square of $\chi_\nu ^2 \approx  1.13$ (6.4 keV) and $\chi_\nu ^2 \approx  1.17$ (6.7 keV). However, the small difference in the value of $\chi_\nu ^2$ is not significant enough to choose one line energy of over the other, which can also be seen from the small difference in Bayesian information criteria $\Delta \rm{BIC} \approx 0.51$, where $\rm{BIC}=\chi^2-k\ln{n}$ and $k$ is the number of model parameters and $n$ the number of data points.
For both line energies, we obtain a rather narrow and similar posterior distribution around the most probable values for all the fitted parameters as seen in Fig. \ref{fig:MCMC_64} and \ref{fig:MCMC_67}. Namely, we can constrain the $M_{\rm BH}$ to be around 3-6 million solar masses, $h_{\rm LP}$ to be between 10-20 $R_g$ above the BH and $R$ to be around 0.5-0.6. (Note that the best fitting parameters shown in Fig. \ref{fig:Best_fit} are not the same as the most probable parameters given by the 50th percentile in Fig. \ref{fig:swift_MCMC} since the probability distributions are not symmetric.)
The degeneracy existing between $M_{\rm BH}$ and $h_{\rm LP}$ discussed previously is reflected as the linear correlation in the $h_{\rm LP} - M_{\rm BH}$ panel. The low $R$ value means that the flux of the reflection component is around 50-60\% of the continuum flux in this energy range. 
We note that \citet{Kara16} used the calibrated correlation between the $M_{\rm BH}$ and lag time of Seyfert galaxies and estimated a $M_{\rm BH}$ value which is of the same order of magnitude but slightly smaller than the $M_{\rm BH}$ value we obtained. This discrepancy is consistent with our prediction that the scaling calibration between $M_{\rm BH}$ and lag time for the two disk geometries should not be the same, since the super-Eddington funnel geometry produces a shorter lag compared to a thin disk for the same $M_{\rm BH}$ and $h_{\rm LP}$, although other complications can also contribute to this discrepancy, e.g., $R$ might not be constant over this energy range.

As a comparison, we also fit a standard thin disk and show the results in Fig. \ref{fig:energy-lag_swift_thin_64} and \ref{fig:energy-lag_swift_thin_67}. We see that thin disk viewed at a large inclination ($i=70^\circ$ and $i=65^\circ$ for the two rest frame Fe line energies) can also fit the data with $\chi_\nu ^2 \approx 1.25$ for both.
The thin disk models give a smaller $M_{\rm BH}=(2-4)\times 10^6 M_\odot$, a similar $ h_{\rm LP} = 9-19 R_g$, and a slightly larger $R=0.65-0.95$ as seen in Fig. \ref{fig:MCMC_thin_64} and \ref{fig:MCMC_thin_67}.
The thin disk fits give BIC values larger by a factor of 2-3, which means that the super-Eddington disk model is slightly preferred over the thin disk model.
This large disk inclination angle may induce a problem when explaining the jet being aligned with the line of sight since it is usually believed that the orbital angular momentum of the innermost accretion disk, the jet direction, and the spin axis of the BH are aligned  \citep[][]{Bardeen75,Stone2012}
\citep[also see Sec. 4 and references therein of the review paper][]{Dai21a}.

We do not attempt to directly fit the Fe K$\alpha$ spectral line profile observed in Swift J1644, given line flux above the continuum is weak. 
However, if we use the best-fit parameters obtained from fitting the lag spectra to calculate a line profile, then the thin disk Fe line would have a very small portion of the emission between the observed energies 7.5-8.5 keV (less than 20$\%$) and almost none reaching 8.5 keV. The super-Eddington disk Fe line, on the other hand, has a higher portion of the spectra in the observed energy range ($\sim 30 \%$) and can reach energies above 8.5 keV. The line profiles from both disk models, however, appear too broad compared to the observed line width. This could be due to factors such as: 1) Some regions along the photosphere are over/under ionized and should not contribute to the Fe line production. 2) For the super-Eddington disk geometry, the coronal photons can have multiple reflections in the funnel and multiple scatterings in the wind. Including the latter into the calculation can introduce additional time delays and energy shifts, which will change the fitting of parameters. We plan to extend our calculations to include these components in a future study.

Lastly, we briefly mention that the observed isotropic X-ray luminosity of Swift J1644 at the peak is $L_X\approx 10^{47} \rm{\ erg \ s^{-1}}$, which is usually interpreted as the jet emission is highly beamed inside the relativistic jet. However, the low dilution factor obtained from our fits suggests that if the corona production is linked to the jet, then the base of the jet ($\sim 10 R_g$) has not accelerated to a relativistic speed yet, consistent with the proposal in \citet{Kara16}.  We also note that there exists another theoretical explanation of the observed reverberation lag \citep{Lu2017}, where the corona is assumed to be far away from the BH ($\approx 100 R_g$) and moving at relativistic speed together with the jet. 
Since we assume a stationary corona and do not include coronal multiple Compton scatterings in our model, we cannot test this hypothesis.

\section{Summary and Discussions}\label{sec:discussion}

In this study, we extend the technique of X-ray reverberation, which has been extensively used to probe geometrically thin accretion disks, to the regime of super-Eddington accretion. The main results are summarized as below:
\begin{itemize}
    \item The coronal photons are reflected inside a funnel surrounded by the optically thick wind launched by the super-Eddington accretion disk. As a result, the coronal emission and its reflection are visible only when the observer looks into the funnel. Due to the narrow funnel opening angle, the reflected photons have shorter lags compared to those from the thin disk geometry.  Also, the Fe K$\alpha$ fluorescent line induced by coronal irradiation has a large blueshift, since the wind flows almost radially towards the observer with a terminal speed of few$\times0.1c$. 
    \item Another consequence of the narrow funnel geometry is that the frequency-dependent Fe K lag behaves almost like a step function near the first minimum (i.e., transiting from being constant at low frequencies to dropping abruptly). As a result, the shape of the energy-lag spectrum is independent of the $M_{\rm BH}$, while the magnitude of the lag scales linearly with the $M_{\rm BH}$. 
    %\item Since the Fe line lag-energy spectrum typically mimics the Fe line emission spectrum to first order, the results of our previous paper \citetalias{Thomsen2019} focusing on the Fe line profile can be also carried over to describe the lag-energy spectrum. Therefore, a largely blueshifted and less skewed Fe K lag-energy spectrum indicates a super-Eddington funnel/wind geometry.
    \item We propose that the funnel/wind reflection geometry produced in super-Eddington accretion systems can be identified using the following Fe K$\alpha$ line characteristics: 1) In the lag-frequency spectrum:  a drastic decrease from constant lag to the first minimum; 2) In the lag-energy spectrum: the spectrum profiles in two low-frequency, narrow bands remain almost the same; 3) In the spectral line / lag-energy spectrum: a large blueshift and less skewed profile, especially when the viewing angle can be independently constrained to be small.
    
    \item We have fitted the observed lag-frequency and lag-energy spectra of the jetted TDE Swift J1644 with our modeling. The MCMC fitting shows that the super-Eddington model fits the observed lag spectra slightly better than any of the thin disk configurations. The best fit using the super-Eddington model gives $M_{\rm BH}= (5-6)\times10^6 \ M_\odot$ and $h_{\rm LP}=10-13 \ R_g$. 
\end{itemize}

In summary, in this work we have shown that the X-ray reverberation lags produced from super-Eddington accretion flow have morphologically different signatures compared to thin accretion disk lags. Such signatures can be used to identify super-Eddington accretion systems around black holes. Furthermore, we have demonstrated that modeling the lag spectrum with a realistic super-Eddington disk structure is necessary for precisely constraining physical parameters from observations. In the future, we plan to further explore how X-ray reverberation signals can help us constrain the funnel geometry and wind acceleration profile, which are likely linked to the accretion rate and the BH spin. We also plan to carry out more detailed radiative transfers studies to include the multiple scattering of photons inside the super-Eddington funnel and the accretion flow. These theoretical studies, coupled with observations from current and next-generation X-ray instruments such as \textit{Athena} and \textit{XRISM}, will shed light on how black holes grow at super-Eddington rates and give feedback to their host galaxies by launching powerful winds.

\acknowledgments{We are grateful to Javier García, Enrico Ramirez-Ruiz and Dan Wilkins for useful discussions. We also thank the anonymous referee for constructive comments. LT and LD acknowledge the support from the Hong
Kong government through the GRF Grants (HKU17305920 and HKU27305119) and the HKU Seed Fund 104005595. Some of the simulations carried out for this project were performed on the HPC computing facilities offered by ITS at HKU and the Tianhe-2 supercluster.
EK acknowledges support from NASA~ADAP grant 80NSSC17K0515.
CSR thanks the UK Science and Technology Facilities Council (STFC) for support under the Consolidated Grant ST/S000623/1, as well as the European Research Council for support under the European Union’s Horizon 2020 research and innovation program (grant 834203)}.

\bibliographystyle{yahapj}
\bibliography{references}

\appendix
\section{Appendix}

\subsection{General Relativistic Ray-Tracing Code} \label{sec:GR_Ray-tracing}
We employ the same GR ray-tracing code as in \citetalias{Thomsen2019} \citep[based on \citealt{Dai2012} and equations from][]{Fuerst04} to calculate the photon trajectory from the corona to the emitting gas and from a faraway stationary observer to the emitting gas. The code uses Boyer-Lindquist coordinates, so the space-time around a rotating (Kerr) BH is given by the line element, $ds^2$, which in geometric units ($G=M=c=1$) is:
\begin{equation} \label{Eq:Kerrmetric}
ds^2 =-\Bigg(1-\frac{2r}{\Sigma}\Bigg)~dt^2 - \frac{4ar  \mathrm{sin^2}\theta}{\Sigma}~dt~d\phi + \frac{\Sigma}{\Delta}~dr^2 + \Sigma~d\theta^2 + \Bigg(r^2 + a^2 + \frac{2a^2r \mathrm{sin^2}\theta}{\Sigma}\Bigg) \mathrm{sin^2}\theta ~d\phi^2,
\end{equation}
where ($t, r,\theta, \phi$) is the Boyer-Lindquist spherical coordinates, $a$ is the dimensionless spin parameter and $\Delta=r^2-2\ r+a^2$ and $\Sigma=r^2+a^2\ \mathrm{cos^2}\theta$. 
In this code, we calculate the geodesic of particles by evolving the six variables $t, r, \theta, \phi, p_r, p_\theta$ using the following equations:
\begin{align}
    & p_t = -E        &&  \dot{t} = E + \frac{2 r (r^2+a^2)E-2 a L}{\Sigma \Delta} \nonumber \\
    & p_r = \frac{\Sigma}{\Delta} \dot{r}   &  &\dot{p}_r = \frac{(r-1)\big((r^2+a^2)H- \kappa \big) + r H  \Delta + 2 r(r^2 + a^2)E^2-2 a E L}{\Sigma  \Delta} - \frac{2 p_r ^ 2(r-1)}{\Sigma} \nonumber\\
    & p_\theta = \Sigma \dot{\theta}  &    &\dot{p}_\theta = \frac{\rm sin\theta  \rm cos\theta}{\Sigma } \Bigg(\frac{L^2}{\rm sin^4\theta}- a^2(E^2+H) \Bigg) \nonumber\\
    & p_\phi = L && \dot{\phi} = \frac{2 a r E + (\Sigma-2 r) L/ \mathrm{sin^2}\theta}{\Sigma  \Delta} \nonumber\\
\end{align}
Here, $p_t$ and $p_\phi$ are constants of motion representing the conservation of energy, $E$, and angular momentum around the spin axis, $L$. Furthermore, we have two additional constants: $H$ is the two times the Hamiltonian (which is 0 for photons and -1 for massive test particles) and $\kappa=Q+L^2+ a^2(E^2+H)$, where $Q$ is the Carter's constant given by $ Q=p_\theta^2 -a^2E\mathrm{cos^2}\theta + L^2\mathrm{cot^2}\theta$.
The six variables are evolved with a standard 4th order Runge-Kutta integrator, which allows us to calculate the photon energy shift, $g$, from the frame of the emitting gas to the observer. We use Louiville's Theorem, which states $I_{\nu}/{\nu^3}$ is conserved along the ray, to relate the observed specific intensity to the emitted intensity \citep{Cunnigham75}:
\begin{equation} \label{Eq:energy_iobs}
    g=\frac{E_{\rm obs}}{E_{\rm emit}}=\frac{(p_{\mu} u^{\mu})_{\rm obs}}{(p_{\mu}  u^\mu)_{\rm emit}}, \ \  \ \ I_{\nu,\rm obs}=g^3 I_{\nu,\rm emit}  \ \ \textrm{and}  \ \ I_{\rm obs}= \int I_{\nu,\rm obs} d\nu_{\rm obs}= \int I_{\nu, emit} g^3 d(g \nu_{\rm emit})  \propto g^4 I_{\rm emit} \propto g^4 \epsilon,
\end{equation}
where $I_{\rm emit}$ is the emitted intensity and it is proportional to the emissivity profile, $\epsilon$, calculated in Sec. \ref{sec:Emissivity}.

\subsection{Local Isotropic Irradiation in GR} \label{Appendix:Isotrpoic}

We adopt the conventional lamppost model which assumes the corona is a point source and it radiates isotropically in its own local frame. Note, this does not mean that the corona is radiating isotropically as seen by a faraway observer in the Boyer-Lindquist (BL) frame.

Isotropic radiation by a point source in the Minkowski space-time (denoted with hat) can be achieved by emitting photons in equally spaced solid angles of $d\hat{\Omega}$: 
\begin{equation}
d\hat{\Omega }=\mathrm{sin}(\alpha)d\alpha d\beta =-d(\mathrm{cos}(\alpha))d\beta,
\end{equation}
where $\alpha$ is the declination angle from the polar axis and $\beta$ is the azimuthal angle.
Therefore, radiating isotropically in Minkowski space can be approximated by sampling random values of $\mathrm{cos}(\alpha)\in [-1,1]$ and $\beta \in [-\pi, \pi]$.
If a photon with energy $E_0$ is emitted with angles $\alpha$ and $\beta$, its energy-momentum 4-vector is found by the standard spherical projection of the solid angles to the Minkowski coordinate tetrad ($\hat{t}, \hat{x}, \hat{y}, \hat{z}$). The projection gives:
\begin{equation} \label{momentum}
\hat{p}=\dot{x}=\Big(p^{\hat{t}}, p^{\hat{x}}, p^{\hat{y}}, p^{\hat{z}} \Big)= E_0\Big(1, \rm sin(\alpha)sin(\beta), sin(\alpha)cos(\beta), cos(\alpha)\Big),
\end{equation}
where we use $p^\mu=\frac{dx^\mu}{d\lambda}=\dot{x^\mu}$ for massless particles.

According to general relativity, one can shift reference frame to the local reference frame (e.g. the corona), so the metric reduces to that of Minkowski space: 
\begin{equation}\label{eq:app:eta}
\eta_{\hat{\alpha} \hat{\beta}} = e_{\hat{\alpha}}^\mu e_{\hat{\beta}}^\nu g_{\mu \nu}= \rm{diag}(-1,1,1,1).
\end{equation}
The transformation matrix is sometimes referred to as 'vierbein matrix' ($e^{\mu}_{\alpha}$) since it consists of four-vectors ('legs'), which makes up the $(4\times 4)$ transformation matrix.
Each of the four legs is normalized and orthogonal to the others. The transformation matrix for arbitrary coronal point-sources orbiting with the four-velocity is $\vec{u}=u^t(1,0,0,\Omega)$ around a Kerr BH is derived in Appendix\ref{Appendix:Vierbein}. The momentum transformation between the local reference frame of a rotating observer and the BL coordinate frame is given by:
\begin{equation} \label{transformation2}
p^\mu=\frac{d x^\mu}{d\lambda}=\dot{x^\mu} = e_{\hat{\mu}}^\mu \hat{p^\mu},
\end{equation}
where $\hat{p^\mu}$ is the four-momentum expressed in the local tetrad basis and $p^\mu$ is the transformed four-momentum expressed in the global BL coordinate frame. We use Eq. \ref{transformation2} to apply the initial conditions of isotropic emission (Eq. \ref{eq:emissivity:stationary}) in the ray-tracing code (Sec. \ref{sec:GR_Ray-tracing}) to estimate $N(\rho,d\rho)$ and $g_{\rm LP}$.
From this relation, we can obtain the three constants of motion (E, L, Q), which can be used to evolve the 6 equations of motion (see Section \ref{Appendix:Vierbein}).

\subsubsection{Vierbein Transformation Matrix} \label{Appendix:Vierbein}

The transformation to local frame in GR is the vierbein transformation matrix, which is a $(4\times 4)$ matrix consisting of four 'legs' $\Big( e_{\hat{t}}, e_{\hat{x}},e_{\hat{y}},e_{\hat{z}} \Big)$, and can be found using the following transformation equations from Eq. \ref{eq:app:eta}. A hatted vector means as measured in its local reference frame and unhatted parameters are measured in the 'global' Boyer-Lindquist coordinates.

The normal convention is to set the time-like, first leg to be the 4-velocity since it automatically fulfills the equation \ref{eq:app:eta} due to the velocity normalization criteria $g_{\mu \nu} u^\mu u^\nu =-1$.
In the Boyer-Lindquist coordinate frame, the velocity of a rotating coronal source is given by $\vec{u}=u^t(1,0,0,\Omega)$. 
From the velocity normalization, one finds $u^t=\sqrt{\frac{-1}{g_{tt}+2\Omega g_{t\phi}+\Omega^2 g_{\phi \phi}}}$. Therefore, the first of the four legs of the vierbein transformation matrix is:
\begin{equation} \label{eq:ZAMO_t}
e_{\hat{t}}^\mu=\Bigg(\sqrt{\frac{-1}{g_{tt}+2\Omega g_{t\phi}+\Omega^2 g_{\phi\phi}}},0,0,\sqrt{\frac{-\Omega^2}{g_{tt}+2\Omega g_{t\phi}+\Omega^2 g_{\phi\phi}}} \Bigg).
\end{equation}

The two next legs can easily be calculated by noting that the local basis ($\hat{z}, \hat{y}$) is parallel to the Boyer-Lindquist basis vector ($\vec{r},\vec{\theta}$). Since there is no cross term in $g_{\mu \nu}$ for $(\vec{r}, \vec{\theta})$, the two legs of the vierbein transformation matrix can be calculated using the normalization condition given in equation \ref{eq:app:eta}, so: 
\begin{equation}\label{eq:ZAMO_r}
e^\mu _{\hat{z}}=\Bigg(0,\sqrt{\frac{1}{g_{rr}}},0,0\Bigg)=\Bigg(0,\sqrt{\frac{\Delta}{\Sigma}},0,0\Bigg),
\end{equation}
\begin{equation}\label{eq:ZAMO_theta}
e^\mu _{\hat{y}}=\Bigg(0,0,\sqrt{\frac{1}{g_{\theta \theta}}},0\Bigg)=\Bigg(0,0,\sqrt{\frac{1}{\Sigma}},0\Bigg).
\end{equation}

The last leg in the vierbein transformation matrix is ($e^\mu _{\hat{x}}$), which needs to account for the cross term in the metric element $g_{t\phi}$, so a simple ansatz is that the last leg has the following form:
\begin{equation}
e^\mu _{\hat{x}}=(e^t _{\hat{x}},0,0,e^\phi _{\hat{x}}).
\end{equation}
Furthermore, we use that it has to be orthogonal to the other transformation bases (legs), which yields:
\begin{align}
    & g_{\mu \nu} e^\mu _{\hat{x}} e^\nu _{\hat{t}}=0 \nonumber\\
    & g_{tt} e^t _{\hat{x}} e^t _{\hat{t}} + g_{t \phi} e^t _{\hat{x}} e^{\phi} _{\hat{t}} + g_{\phi t} e^{\phi} _{\hat{x}} e^t _{\hat{t}} + g_{\phi \phi} e^{\phi} _{\hat{x}} e^{\phi} _{\hat{t}} = \nonumber\\
    & e^t _{\hat{x}} \Big(g_{tt} e^t _{\hat{t}} + g_{t \phi} e^{\phi} _{\hat{t}} \Big) + e^{\phi} _{\hat{x}} \Big(g_{\phi t}e^t _{\hat{t}} + g_{\phi \phi}  e^{\phi} _{\hat{t}} \Big)=0.
\end{align}
So, there exist the following simple relation between the two components:
\begin{equation} \label{compare}
e^{\phi} _{\hat{x}}=\frac{-\Big(g_{tt} e^t _{\hat{t}} + g_{t \phi} e^{\phi} _{\hat{t}} \Big)}{\Big(g_{\phi t}e^t _{\hat{t}} + g_{\phi \phi}  e^{\phi} _{\hat{t}} \Big)} e^{t}_{\hat{x}}=\frac{-\Big(g_{tt} + \Omega g_{t\phi} \Big)}{\Big(g_{t\phi}+\Omega g_{\phi \phi} \Big)} e^{t}_{\hat{x}}=\frac{-g_{tt}-\Omega g_{t\phi}}{g_{\phi \phi}(\Omega-\omega)} e^{t}_{\hat{x}},
\end{equation}
where we have used that $e^\phi _{\hat{t}}=\Omega e^t _{\hat{t}}$, and we have introduced the frame dragging velocity $\omega=\frac{-g_{t\phi}}{g_{\phi\phi}}$.
The transformation (normalization) condition from equation \ref{eq:app:eta} gives the following expression:
\begin{align}
& g_{tt} e^t _{\hat{x}} e^t _{\hat{x}} + 2 g_{t \phi} e^t _{\hat{x}} e^{\phi} _{\hat{x}} + g_{\phi \phi} e^{\phi} _{\hat{x}} e^{\phi} _{\hat{x}}= \nonumber\\ 
& \big(e^t _{\hat{x}} \big)^2 \bigg(g_{tt}-2g_{t\phi}\Big(\frac{g_{tt}+\Omega g_{t\phi}}{g_{\phi \phi}(\Omega-\omega)}\Big)+ g_{\phi \phi} \Big(\frac{g_{tt}+\Omega g_{t\phi}}{g_{\phi \phi}(\Omega-\omega)}\Big)^2 \bigg)=1.
\end{align}
Next, the basis (leg) component $e^t_{\hat{x}}$ can be isolated, and from Eq. \ref{compare}, we can quickly find the expression for the other basis (leg) component $e^{\phi}_{\hat{x}}$. The full expression for the last leg of the vierbein transformation matrix is:
\begin{align} \label{eq:ZAMO_phi}
& e^\mu_{\hat{x}}=\Bigg(\frac{\Omega-\omega}{\sqrt{g_{tt}\big(\Omega-\omega\big)^2-2g_{t\phi} (\Omega-\omega\big) \Big(\frac{g_{tt}+\Omega g_{t\phi}}{g_{\phi\phi}}\Big) + \frac{\big(g_{tt}+\Omega g_{t\phi}\big)^2}{g_{\phi\phi}}}},0, \nonumber \\
& 0,\frac{-\big(g_{tt}+\Omega g_{t\phi}\big)}{\sqrt{ g_{tt} g_{\phi\phi}^2 \big(\Omega-\omega\big)^2-2g_{t\phi} g_{\phi\phi} (\Omega-\omega\big) \Big(g_{tt}+\Omega g_{t\phi} \Big) + g_{\phi\phi}\big(g_{tt}+\Omega g_{t\phi}\big)^2}}\Bigg).
\end{align}

The transformation simplifies significantly for a stationary observer with $\Omega=0$ and an observer rotating with the frame-dragging speed $\Omega=-\frac{g_{\phi t}}{g{\phi \phi}}=\omega$, which is known as a ZAMO (Zero-Angular-Momentum-Observer).

\subsubsection{Stationary observer} \label{Appendix:stationary}
 A stationary observer with $\Omega=0$ will have the following simplified expressions of $e^\mu _{\hat{t}}$ and $e^\mu _{\hat{x}}$:
\begin{align}
    &e_{\hat{t}}^\mu=\Bigg(\sqrt{\frac{-1}{g_{tt}}},0,0,0 \Bigg)_{\rm STATIONARY} \\
    & e^\mu_{\hat{x}}=\Bigg(\frac{1}{\sqrt{g_{\phi \phi} \big(\frac{g_{t t}}{g_{t \phi}}\big)^2-g_{tt}}},0,0,\frac{-1}{\sqrt{g_{\phi \phi} -\frac{(g_{t \phi})^2}{g_{tt}}}}\Bigg)_{\rm STATIONARY}.
\end{align}
From equation \ref{transformation2}, the initial four momentum of a photon emitted by a stationary corona in the Boyer-Lindquist coordinates can be expressed as:
\begin{align} \label{eq:emissivity:stationary}
& \dot{t}_{\rm STATIONARY}=p^t =e_{\hat{t}} ^t p^{\hat{t}}+e_{\hat{x}} ^t p^{\hat{x}} = \frac{-E_0}{\sqrt{-g_{tt}}}+\frac{E_0 \mathrm{sin}(\alpha)\mathrm{cos}(\beta)}{\sqrt{g_{\phi \phi} \big(\frac{g_{t t}}{g_{t \phi}}\big)^2-g_{tt}}} \nonumber\\
&\dot{r}_{\rm STATIONARY}=p^r =e_{\hat{z}} ^r p^{\hat{z}}=E_0 \mathrm{cos}(\alpha)\sqrt{\frac{\Delta}{\Sigma}} \nonumber\\
& \dot{\theta}_{\rm STATIONARY}=p^\theta=e_{\hat{y}} ^\theta p^{\hat{y}}=\frac{E_0 \mathrm{sin}(\alpha)\mathrm{sin}(\beta)}{\sqrt{\Sigma}}\nonumber\\
& \dot{\phi}_{\rm STATIONARY}=p^\phi =e_{\hat{x}} ^\phi p^{\hat{x}}=\frac{-E_0 \mathrm{sin}(\alpha)\mathrm{cos}(\beta)}{\sqrt{g_{\phi \phi} -\frac{(g_{t \phi})^2}{g_{tt}}}}.
\end{align}

\subsubsection{ZAMO} \label{App:ZAMO}
In a Kerr space-time, a stationary observer is an observer that is co-rotating with the space-time geometry of the BH since objects within the ergosphere cannot appear stationary to a truly static observer far away \citep{Bardeen72}. This observer, rotating with the frame-dragging velocity $d\phi/d\tau=\Omega=\omega=-g_{t \phi}/g_{\phi \phi}$, is known as a ZAMO \citep{Krawczynski2017} and it is the standard observer in GR.
The ZAMO observer is characterized by having zero angular momentum and its position is fixed at constant $r$ and $\theta$, so $p_\phi=L_z=\dot{r}=\dot{\theta}=0$.
For a ZAMO, the two legs of the transformation matrix ($\hat{t}, \hat{x}$) simplifies to: 
\begin{align} \label{eq:ZAMO1_tphi}
& e_{\hat{t}}^\mu=\Bigg(\sqrt{\frac{-1}{g_{tt}- \frac{g_{t\phi}^2}{g_{\phi \phi}}}},0,0,\sqrt{\frac{-1}{g_{tt} \frac{g_{\phi \phi}^2}{g_{t\phi}^2}- g_{\phi \phi}}} \Bigg)_{\rm ZAMO} \\
& e_{\hat{x}}^\mu = \Bigg(0,0,0,\frac{-1}{\sqrt{ g_{\phi\phi}}}\Bigg)_{\rm ZAMO}.
\end{align}
This simple transformation of $e_{\hat{x}}^\mu$, for a ZAMO, means a particle with velocity $\dot{\phi}=0$ in the BL frame will keep having zero velocity in the ZAMO-frame, which is an additional advantage of such an observer. If the local reference frame of an object differs from that of the ZAMO (other velocity profile), then one can always get the local reference frame by adding the Lorentz-factor, which takes the relative motion between the ZAMO and local reference frame of the object into account \citep{Bardeen72}. 
A ZAMO observer has the 3-velocity $v_\phi=v_r=v_\theta=0$, so the relative velocity is found by transforming the 4-velocity of the particle $U^{\mu}$ into the ZAMO frame  with the following equation:
\begin{equation} \label{eq:3vel}
    v_{\hat{i}}=\frac{g_{\mu \nu} U^\mu e_{\hat{i}}^\nu }{g_{\mu \nu} U^\mu e_{\hat{t}}^\nu}.
\end{equation}

\subsection{Relativistic proper Area} \label{App:Area}
We find the GR proper area of surface elements by performing the Jacobian. Our super-Eddington disk structure has been averaged in the azimuthal ($\phi$) direction, so it is axis-symmetric in the z-direction. Therefore, we are looking for a mapping F that maps the photosphere from ($r, \theta, \phi$) to $(\rho, \phi)$:
\begin{equation}
\Big(\rho, \phi \Big) \frac{F}{\mapsto } \Big(r, \theta, \phi \Big) = \Big(r(\rho), \theta(\rho), \phi \Big).
\end{equation}
The transformation is given by the metric tensor:
\begin{equation}
\begin{aligned}
h=
	\begin{bmatrix}
	\big< \Sigma_k \lambda_{k \rho} \lambda_{k \rho}\big> & \big< \Sigma_k\lambda_{k \rho} \lambda_{k\phi}\big> \\
	\big< \Sigma_k \lambda_{k\phi} \lambda_{k\rho} \big> & \big< \Sigma_k \lambda_{k\phi} \lambda_{k\phi}\big>
	\end{bmatrix}
=
	\begin{bmatrix}
	g_{rr} \big(\frac{\partial r}{\partial \rho}\big)^2 + g_{\theta \theta} \big(\frac{\partial \theta(\rho)}{\partial \rho}\big)^2  & 0\\
	0 & g_{\phi \phi} \frac{\partial \phi}{\partial \phi}
	\end{bmatrix},
\end{aligned}
\end{equation}
where $\lambda_{ij}=\frac{\partial F_i}{\partial u_j}$ is the Jacobian, and $F$ is the mapping from $\big(r(\rho), \theta(\rho), \phi \big)$ to $u=(\rho, \phi)$. 
Here, $<*>$ denotes the inner product under the metric, and $\Sigma_k$ is the summation over all indices $k$. 
The proper area as seen by a stationary observer is:
\begin{equation}
A_{\rm stationary}=\sqrt{\mathrm{det}(h)}d\rho d\phi,
\end{equation}
where $\mathrm{det}(h)$ is the determinant of h.
The subscript "stationary" means the proper area in the frame of a ZAMO observer. In order to get the area as seen by the super-Eddington disk, we need to add the Lorentz factor between the ZAMO and disk elements using Eq. \ref{eq:3vel}

\end{document}